\documentclass[runningheads]{llncs}

\usepackage[T1]{fontenc}  
\usepackage[english]{babel}  
\usepackage{csquotes}
\usepackage{microtype}  
\usepackage{fnpct}  
\usepackage{orcidlink}

\usepackage{booktabs}
\usepackage{graphicx}  
\graphicspath{{./figs/}}  

\usepackage{mathtools}

\usepackage{amssymb}

\usepackage{algorithm}
\usepackage{algpseudocode}

\renewcommand*{\emptyset}{\varnothing}  

\DeclareSymbolFont{bbold}{U}{bbold}{m}{n}
\DeclareSymbolFontAlphabet{\mathbbold}{bbold}
\newcommand*{\ind}{\mathbbold{1}}  

\newcommand*{\phanrel}{\mathrel{\phantom{=}}}

\newcommand*{\Reals}{\mathbb{R}}

\DeclareMathOperator{\E}{\mathbb{E}}
\DeclareMathOperator*{\argmax}{arg\,max}

\DeclarePairedDelimiterX\set[1]\lbrace\rbrace{\mkern1.5mu\def\suchthat{\;\delimsize|\;}#1\mkern1.5mu}

\usepackage{hyperref} 
\usepackage{ellipsis}  

\newcommand{\bn}{\bigskip\noindent}
\newcommand{\mn}{\medskip\noindent}

\newcommand*{\NW}{\mathcal{NW}}

\newcommand{\EFo}{EF1}
\newcommand{\WEF}{WEF}
\newcommand{\EREF}{EREF}

\newcommand*{\OPT}{A^\mathrm{OPT}}
\newcommand*{\DEC}{A^\mathrm{DEC}}
\newcommand*{\CEN}{A^\mathrm{CEN}}
\newcommand*{\NUL}{A^\mathrm{\emptyset}}

\begin{document}

\title{Decentralized Fair Division}

\author{
    Joel Miller\orcidlink{0009-0003-3817-9042} \and
    Rishi Advani\orcidlink{0000-0002-5522-0401} \and
    Chris Kanich\orcidlink{0000-0002-3836-2168} \and
    Lenore D Zuck\orcidlink{0000-0003-3613-1208} \and
    Ian A. Kash\orcidlink{0000-0002-7826-8555}
    }

\authorrunning{Miller et al.}

\institute{University of Illinois at Chicago, Chicago IL 60607, USA\\
\email{\{jmill54,radvani2,ckanich,zuck,iankash\}@uic.edu}}

\maketitle

\begin{abstract}
Fair division is typically framed from a centralized perspective. However, in practice resource allocation often occurs via decentralized networks. We study a \emph{decentralized} variant of fair division inspired by altruistic dynamics observed in behavioral economics and other practical settings. We develop an approach for decentralized fair division and compare it with a centralized approach with respect to fairness and social welfare guarantees. Our decentralized model can be seen as a relaxation of previous models of sequential exchange, in light of impossibility results concerning the inability of those models to achieve desirable outcomes. 
We find that the two models of resource allocation offer contrasting fairness and social welfare guarantees, and map out how these guarantees depend on valuations and other model parameters. We further show conditions under which a mix of the two approaches outperforms either approach in isolation.
Despite the simplicity of our decentralized model, we show that under appropriate conditions it can ensure high-quality allocative decisions in an efficient fashion.

\end{abstract}

\section{Introduction}\label{sec:intro}
Fair division research has made immense strides in recent years and carries serious potential to improve the fairness of resource allocations in many practical settings. Most of this research is framed from an idealized \emph{centralized} perspective: a single decision maker, who has access to all relevant information, decides on the resource allocation and enforces it with omnipotence. 

Real-life allocative scenarios do not always exhibit these features. Instead, a large body of literature from 
economics~\cite{alderman2002do,conning2002community,hussam2022targeting,sobel2005interdependent}, sociology~\cite{calomiris1998role,bouman1995rosca,matthewman2020sociology,miao2021responding}, and anthropology~\cite{yan2012gift,spade2020solidarity,kneale2023mobilising,chevee2022mutual} has shown that \textit{decentralized} networks of cooperating agents often emerge to tackle allocative challenges. Decentralized networks have been effective tools for distributing food~\cite{Hanson_Coupal_Grace_Jesch_Lockhart_Volpe_2024,Lofton_Kersten_Simonovich_Martin_2022}, personal protective equipment~\cite{chevee2022mutual},  and disaster relief services~\cite{kenney2019solidarity} -- in fact, ``Decentralised\ldots disaster relief efforts have arisen after nearly every major natural disaster in the United States since Katrina''~\cite{firth2022disaster}. 

These decentralized schemes have had large real-world impacts but existing fair division research, which focuses on centralized algorithms, feels ill-equipped to explain why these networks have been successful. This is partly because these networks operate in ways that are not easily modeled by existing tools in fair division research. Indeed, our work hinges on three assumptions which are not common in the rest of the literature.
\begin{itemize}
    \item \textit{Altruism.} Much economics literature assumes that individuals are selfish and utility-seeking. While agents in fair division research aren't typically assumed to strategize for the sake of maximum utility, it is common to assume that fairness looks like every agent getting \textit{at least} some benchmark level of utility. Assumptions about utility-seeking agents are sensible for studying many types of economic activity, but they cannot easily explain the altruism practically observed in these networks ~\cite{spade2020solidarity,kneale2023mobilising,chevee2022mutual}. Indeed, the literature on dictator games shows that individuals balance their own utilities with overall welfare even when they have unilateral power over the allocative decision at hand ~\cite{engel2011dictator}, and that individuals who are closely linked in a social network exhibit more altruism~\cite{leider2010we}. Hauser \textit{et~al.}~\cite{hauser2019social} show that participants in public goods games can exhibit similar altruistic behavior, including a tendency for well-off\footnote{Within the context of the game.} participants to ``give" to less well-off participants.

    \item \textit{Local knowledge.} Fair division research tends to assume an omniscient central planner (i.e. one who knows valuations perfectly). This assumption makes sense in many contexts, but it seems to not always be true in the decentralized settings we are interested in. Empirical evidence suggests that when decentralized networks do succeed, they succeed partly because individuals are able to take advantage of local information that centralized authorities lack access to ~\cite{alderman2002do,conning2002community,hussam2022targeting}. Literature from Computer Science~\cite{schechtman2025discretion} and the Health Sciences~\cite{meehl1954clinical} has shown that human decision makers often use local knowledge to improve upon algorithmic recommendations. 

    \item \textit{Local computational constraints.}     
    While individuals may have a leg up in terms of information, they are also cognitively constrained compared to their centralized counterparts. This can result in local decisions about good allocation that could be considered sub-optimal from a global perspective ~\cite{ortiz2012decisions,dupas2022measuring}. Indeed, recent human studies on fair division suggest that increased fairness comes at the cost of increased cognitive burden ~\cite{hosseini2025epistemic}.
    
\end{itemize}

Thus, our work differs from the main body of literature in its focus on three important assumptions -- \textit{altruism}, \textit{local knowledge}, and \textit{local computational constraints}. While it would be nice to share more modeling choices with the rest of the literature, we find that these assumptions are essential for cleanly describing decentralized systems.

From a purely descriptive perspective, understanding the strengths and weaknesses of decentralized allocation can help economists and computer scientists evaluate the currently existing systems mentioned above.
Moreover, as mutual aid networks and other community-based forms of organization move into the digital space~\cite{wilson2022mutual,koletsi2021virtually}, a technical understanding of decentralized dynamics will aid any future attempts to incorporate fair division into decentralized digital practices. Yet, despite the importance of  decentralized schemes, little fair division research has studied them. 
In this work we develop a decentralized variant fair division to better understand its unique strengths and weaknesses, and thereby address this gap in the literature. 

Since our assumption of local computational constraints results in a decentralized scheme that is computationally simple, this work may be of independent interest to researchers and practitioners interested in simple approximation algorithms for fair division. Indeed, like all decentralized processes, ours can be simulated on a single computer (provided one has access to the appropriate information). As we will prove, the results of the simulation converge to a the results of a simple linear-time algorithm. The linear-time nature of this algorithm may make it appealing for large problem sizes, even without any assumption of decentralization. Additionally, our results on an imperfect central planner may be of interest to anyone interested in understanding Nash welfare maximization in the presence of missing information. Lastly, the present study can be understood as a continuation of work by Igarashi \textit{et~al.} \cite{igarashi2024reachability} and Ramezani \textit{et~al.} \cite{ramezani2010nash}, who analyzed the reachability of fair or welfare-maximizing allocations through sequential exchanges, without requiring exchanges to be individually rational (but with the requirement that they be locally fairness-increasing or welfare-increasing). In light of the negative results obtained by these authors, we further relax the requirements for intermediate allocations in our decentralized exchange procedure, and ask if interesting allocations are still reachable. Perhaps surprisingly, we show that a model which does not enforce any strict fairness or welfare guarantees on intermediate allocations still can produce desirable outcomes overall.

\subsection{Overview of our Model and Contributions}
We use Nash social welfare as the main desiderata by which to measure our allocative schemes. To that end, our novel model of decentralized fair division is a stochastic processes that gradually evolves allocations according to a simple polynomial-time rule which approximates Nash welfare maximization, via modeling choices derived from the above literature.
While we propose a particular process for concreteness, we show it eventually stops changing.  That is, it reaches an equilibrium (in the dynamical systems sense).  We primarily analyze it through this equilibrium point, so most of our results apply to a larger family of processes that converge to the same equilibrium.  Thus, our results are robust to some of the specific details of this process.

We analyze the unique properties of decentralized networks by contrasting our decentralized scheme against a stylized model of a centralized decision maker, who has the ability to exactly maximize Nash welfare, but only relative to imperfect information about the agents. 

In particular, we instantiate each agent with an \emph{endowment} of extra utility which cannot be traded. In our model, endowments are visible during the decentralized allocation process, where the agents use this information in their approximations of Nash welfare.
The endowments are hidden from the centralized mechanism, which can only maximize Nash welfare with respect to the set of goods up for redistribution. This modeling choice captures the observation that communities hold \textit{local knowledge} that centralized entities have difficulty accessing-- endowments represent of the type of information local agents may have but centralized sources may lack. It is specifically inspired by empirical literature on community-based targeting~\cite{conning2002community}, which investigate schemes to ``\textit{decentralize the responsibility for monitoring poverty... to local administrators who, it is argued, should be able to do so more accurately and cost-effectively than a central government agency}''~\cite{alderman2002do}. In other words, empirical evidence shows cases where central authorities lack information on socio-economic status, while locals possess such knowledge. Thus, endowments -- extra un-tradable utility -- can be thought of as stylized stand-ins for socioeconomic status (SES) levels, which are in some cases more readily known in local communities. 
Alternatively, it can capture the logistical challenges of gathering timely information: in the immediate aftermath of a natural disaster it is a substantial effort to establish how badly each individual was affected, while in contrast this may be easily observed locally.

We capture the observation that individuals act altruistically by letting individual agents make choices that decrease their own utility during the stochastic decentralized process. Lastly, we adopt a simple, computationally efficient update rule for the decentralized model (in contrast to the NP-hardness of maximizing Nash welfare) in order to capture the observation that individuals in decentralized networks are cognitively constrained.

Our main results show that the ratio between the good valuations and endowments determines which scheme provides higher Nash welfare (Theorems \ref{thm:MA approx MNW}--\ref{thm:CPLI approx MNW}). We also identify a natural class of instances where a partial combination of the two mechanisms attains better Nash welfare than either in isolation (Theorem~\ref{thm:inbetween}). 
Numerical simulations fill in the picture painted by this theory. Among other things, they suggest a broader range of scenarios where a combination of mechanisms is best.
We also define a new fairness property, endowment relative envy-freeness (EREF), and establish conditions under which the decentralized mechanism attains it (Theorems \ref{thm:1wayEF deterministic}--\ref{thm:1wayEF probability}). 

\subsection{Related Work}\label{subsec:rel work}
Fair division is a rich research area at the intersection of economics and computer science. For a recent general survey of the field, see \cite{aziz2022algorithmic}. 

Our model of decentralized allocation is an example of dynamic fair division, in the sense that goods are reallocated over time.  While our results focus less on this aspect of decentralized allocation, other recent work has focused on the particular challenges introduced by dynamics~\cite{kash2014no,friedman2017controlled,zeng2020fairness,sinclair2022sequential}. Of particular note, Benad{\`e} \textit{et~al.}~\cite{benade2022dynamic} also consider a model with information restrictions, although of quite a different form than we impose on our centralized process. Their algorithm is given access to limited ordinal information about the agents' valuations, but does not know the exact valuations themselves; our centralized process has access to cardinal valuations, but does not see the latent endowments.

Some fair division research focuses on asymmetric agents from a centralized perspective~\cite{chakraborty2021weighted,chakraborty2021picking,chakraborty2022weighted,babaioff2021competitive,babaioff2023fair,viswanathan2023general,aziz2020polynomial,hajiaghayi2023almost,montanari2024weighted,farhadi2019fair}. In these models, agents are given \emph{weights}, where a higher weight represents a larger entitlement to the pot of goods. Our concept of an endowment is similar to a weight, but here a higher endowment represents a \emph{smaller} entitlement to the pot of goods. 
Our novel \EREF{} criterion can be thought of as a weaker version of the notion of weighted envy-freeness (\WEF{}) from this literature, as \WEF{} seems too strong to be achievable in general in our decentralized setting and potentially even undesirable in some contexts as it would limit the amount of assistance provided to agents with smaller endowments.

Some of our analysis focuses on probabilistic guarantees. A strain of fair division literature also seeks to provide probabilistic guarantees ex-ante with randomized algorithms~\cite{aziz2023best,martin2023best,freeman2020best}. In contrast, we focus on randomized valuations. 

\paragraph{Our decentralized model as an extension of the literature.}

Several papers have explored decentralized models of fair resource allocation via local interactions ~\cite{endriss2003optimal,endriss2006negotiating,chevaleyre2007reaching,chevaleyre2017distributed,beynier2018fairness,gourves2017object}. These papers feature individually rational agents who will only trade goods if it improves their utility, and Endriss \textit{et~al.}~\cite{endriss2003optimal,endriss2006negotiating} introduce payment functions to facilitate such individually rational transactions. In contrast, we do not assume individual rationality because it does not correspond to the empirical observations from the work in economic anthropology discussed above. This also obviates the need for payment functions. 
Lange and Rothe study a model of exchange where agents need not be individually rational, but only from a complexity-theoretic perspective~\cite{lange2019optimizing}.

A few other works have analyzed the reachability of fair or welfare-maximizing allocations through sequential exchanges, without requiring exchanges to be individually rational, but with the condition that intermediate states need to be fair or welfare-increasing. Notably, Igarashi \textit{et~al.}~\cite{igarashi2024reachability} study whether specific \EFo{} allocations are reachable via sequential exchanges when intermediate allocations are also required to be \EFo{}, and they prove several impossibility results\footnote{Namely, when agents have general valuations, \EFo{} states are not reachable in general. If valuations are binary or symmetric, \EFo{} states are reachable for cases with two agents, but not with three or more.}. Similarly, Ramezani and Endriss~\cite{ramezani2010nash} study sequential local exchanges of goods that each increase Nash welfare, and they find that reaching a Nash welfare-maximizing allocation may require a ``local'' exchange that in fact includes every agent. Our model of decentralized good exchange can be seen as a continuation of these two studies: in light of their negative results, we further relax the requirements for intermediate allocations in our decentralized exchange procedure, and ask if interesting allocations are reachable. 

\section{Model and Problem Setup}\label{sec:model}
A \emph{fair division instance} is modeled by a tuple 
$$\langle N = \{1.\dots,n\},~G= \{g_1.\dots,g_m\}, ~(v_i)_{i \in N},~\allowbreak (e_i)_{i \in N}\rangle$$ 
where $N$ is a non-empty set of \emph{agents} and $G$ is a non-empty set of \emph{goods}. For every agent $i\in N$, $v_i\colon G \to \Reals_{\geq 0}$ denotes agent $i$'s \emph{valuation} function mapping a good $g\in G$ into $v_i(g)$, the amount of utility agent $i$ receives from $g$. Each agent also has an \emph{endowment} $e_i \in \Reals_{> 0}$, representing an inalienable amount of utility that agent $i\in N$ will enjoy regardless of which goods they receive. We assume that valuations for sets of goods are additive. For $G' \subseteq G$ and $i\in N$, we use $v_i(G')$ to denote $\sum_{g \in G'} v_i(g)$.

Given a fair division instance, an \emph{allocation} $A = (A_1, \dots, A_n)$ is a partition of $G$ where agent $i$ receives the bundle of goods $A_i$. Let $\mathbf{A}$ represent the set of all possible allocations.

\begin{definition}[Nash welfare and Nash welfare maximization]
The Nash welfare of an allocation $A \in \mathbf{A}$ is
\[\NW(A) = \sum_{i \in N} \log(v_i(A_i) + e_i)\,.\]
An allocation $A$ maximizes Nash welfare  if $A \in \arg\allowbreak \max_{A' \in \mathbf{A}} \NW(A')$. Let $\OPT$ denote an allocation maximizing Nash welfare.
\footnote{
Much other work in fair division, going back to Nash's original bargaining paper~\cite{nash1950bargaining}, defines Nash welfare as the product of utilities. We instead use the sum of logs of utilities. From an optimization perspective, the two definitions are equivalent, since taking the log of the product of utilities gives our definition of Nash welfare and log is monotonic. Work that has used Nash welfare as a measure of social welfare (rather than merely a device for finding fair allocations) has traditionally used the log form we adopt~\cite{kaneko1979nash}. This form naturally connects to the use of logarithmic utility for wealth as a way of capturing risk aversion, as in the classic solution to the St.~Petersburg paradox~\cite{stigler1950development}. 
}
\end{definition}

$\OPT$ has appealing properties such as Pareto optimality and, when all endowments are the same, envy-freeness up to 1 good~\cite{caragiannis2019unreasonable}, which we define below. Aside from that, Nash social welfare is an appealing measure to maximize in its own right because it strikes a balance between utilitarian and egalitarian notions of social welfare.

We use \emph{envy-freeness} as the basis for our primary fairness desiderata. An allocation is envy-free if no agent would prefer to swap their bundle with that of another agent. Envy-freeness is a popular concept in the literature so we adopt it here. Envy-freeness is not always satisfiable\footnote{Consider, for example, a scenario with two agents and one good, where both agents value the good positively.}. One well-studied relaxation is \emph{envy-freeness up to 1 good (\EFo{})}.

\begin{definition}[envy-freeness up to 1 good (\EFo{})]
An allocation $A \in \mathbf{A}$ is \EFo{} if for every pair of agents $i,j \in N$, there exists a good $g\in A_j$ such that $v_i(A_i) \geq v_i(A_j \setminus \set{g})$.
\end{definition}

Notice that envy-freeness is defined with respect only to the goods up for redistribution, and not with respect to endowments. In settings with asymmetric agents (such as agents with different endowments), fairness is commonly understood through the lens of weighted envy-freeness (\WEF{})~\cite{chakraborty2024weighted}. Like envy-freeness, WEF is not always satisfiable. Therefore, we introduce a novel relaxation of \WEF{} based on the idea that given disparate endowments, fairness might mean treating those with lower endowments better. In particular, under an interpretation of endowments as stylized stand-ins for SES levels, is natural to want to guarantee that an agent with a lower endowment should never envy the bundle of an agent with a higher endowment. Even though agents do share goods altruistically, we opt to not let agents share their endowments because decentralized exchange networks tend to focus on a specific class of resources like food or medicine, rather than pursuing a complete redistribution of material wealth~\cite{Lofton_Kersten_Simonovich_Martin_2022,Hanson_Coupal_Grace_Jesch_Lockhart_Volpe_2024,chevee2022mutual}. At the same time, common knowledge of other participants' SES levels tend to influence decision making in these networks.

\begin{definition}[Endowment-relative envy-freeness (\EREF{})]
An allocation $A$ is \EREF{} if for every pair of agents $i,j \in N$,
\[e_i < e_j \implies v_i(A_i) \geq v_i(A_j)\,. \]
\end{definition}

Like all relaxations of fairness notions, \EREF{} is imperfect: namely, it does not give any kind of envy guarantee for agents with higher endowments towards agents with lower endowments. However, the intuition is that since these agents already have higher endowments, it is ``ok enough'' to not give them an envy guarantee. Meanwhile, if $e_i$ is less than $e_j$ (i.e. if agent $i$ already envies agent $j$ as far as endowments are concerned), we can at least guarantee that $i$ that they will not also envy $j$ as far as goods are concerned.  

Neither \EFo{} nor \EREF{} imply the other, since \EREF{} gives a stronger fairness guarantee, but only uni-directionally among any pair of agents. While \EREF{} can be trivially satisfied (assign all goods to an agent with the lowest endowment), the Nash welfare of such an allocation may be sub-optimal. We are interested in allocation procedures that can provide guarantees about both \EREF{} and welfare.

\subsection{Centralized Allocation}
Since the literature indicates that centralized entities lack local knowledge in some scenarios, we model a centralized mechanism which can exactly maximize Nash welfare (an NP-hard computation in general~\cite{ramezani2010nash,nguyen2013survey}) but cannot see the endowments of agents --- thus, it produces an allocation $A$ that maximizes Nash welfare, but only with respect to the goods in $G$. We refer to the \emph{centralized solution} to a fair division instance as an allocation 
\[\CEN \in \argmax_{A \in \mathbf{A}}\sum_{i \in N} \log(v_i(A_i)) \,.\]

The restriction that the centralized allocation cannot access endowments is fairly strict -- however, it is the cleanest way to capture the core trade-offs we are interested in understanding. Moreover, as we show below, $\CEN$ can approximate Nash welfare arbitrarily poorly even when it has access to all but two endowments, so in some sense this restriction does not change our basic results with regards to Nash welfare. 

\subsection{Decentralized Allocation}\label{subsec:decentralized def}

We model decentralized resource allocation as a stochastic, iterative processes that gradually evolves allocations according to a simple polynomial-time rule which approximates Nash welfare maximization. We assume that there exists some starting allocation which is then altered over a series of rounds. Algorithm~\ref{alg:dec} describes a single round of such a procedure.

\begin{algorithm}
\caption{One round of decentralized exchange}\label{alg:dec}
\begin{algorithmic}
\Require Allocation $A$, group size $k$ 
\State Pick a random subset $S \subseteq N$ of size $k$
\For{$i \in S$}
    \For{$g \in A_i$}
        \State Pick an agent $j \in \argmax_{i' \in S} \frac{v_{i'}(g)}{e_{i'}}$
        \State $A_{i} \gets A_i \setminus \{g\}$
        \State $A_j \gets A_j \cup \{g\}$
    \EndFor
\EndFor
\end{algorithmic}
\end{algorithm}

In each round of decentralized exchange, a subset of agents $S$ redistributes their currently held goods among themselves. We interpret the process of choosing $S$ as modeling how various groups in a society might organically meet to exchange goods. The redistribution among members of $S$ occurs via a simple rule: give each good to an agent in $S$ maximizing $v_i(g) / e_i$. This rule is an adaptation of Adams' longstanding model of equity in social exchange~\cite{adams1965inequity,folger2000social} to the context of fair division. It also approximates Nash welfare maximization in the sense that\footnote{We show this equality in Proposition~\ref{prop:decentralized alloc rule}, in the appendix.}
\begin{align}\label{math:dec equivalence}
\argmax_{i \in N} \frac{v_i(g)}{e_i} = \argmax_{i \in N} \log(v_i(g) + e_i) - \log(e_i) \,.
\end{align}
The term $\log(v_i(g) + e_i) - \log(e_i)$ describes the marginal increase to Nash welfare that would occur from giving $g$ to $i$, assuming $i$'s bundle is currently empty. So the allocation rule \emph{approximates} Nash welfare maximization in the sense that it exactly maximizes Nash welfare in the world where $G$ is just the singleton set $\{g\}$. Since empirical evidence shows that participants in resource sharing games prefer allocations that improve welfare to allocations that only improve fairness~\cite{charness2001relative},we model agents as trying to improve Nash welfare rather than some fairness metric. The protocol is decentralized in the sense that only a subset of the network is allowed to interact in each round of the process, and in the sense that computation and communication is reduced to a very simple procedure. 

Algorithm~\ref{alg:dec} models key observations from the literature on decentralized allocation since agents take advantage of local knowledge (the endowments) and act altruistically (an agent can decrease their utility by removing a good from their bundle), but are cognitively constrained to poly-time calculation: the re-allocation rule approximates Nash welfare maximization, but agents can ``make mistakes'' by re-allocating a good in a way that actually decreases Nash welfare. The choice to have agents take endowments into account when distributing goods is a normative one, since arguably one's private utility should not be a factor in deciding an allocation. However, we feel that having agents consider endowments is much more reflective of the real-world systems we are interested in understanding, where allocation does not occur separately from everyday life, but is instead highly situated in the broader context of the community.   

In Appendix~\ref{apdx:dec model explanation}, we bound how quickly
\footnote{Specifically, after $t$ rounds of exchange the probability of convergence is at least $1 - m \biggl(1 - \frac{k^2 - k}{n^2 - n}\biggr)^t$.}
repeated application of Algorithm~\ref{alg:dec} converges to an allocation where
\begin{align}\label{math:dec property}
    g \in A_i \iff i \in \argmax_{j \in N} \frac{v_j(g)}{e_j} \,.
\end{align}
This serves as an equilibrium point in the dynamical systems sense since no further local updates are improvements.
Since all of our theoretical results (except Theorem~\ref{thm:inbetween}) only examine the equilibrium state of this process, these results are insensitive to some of the details of the decentralized procedure, such as the initial allocation, the way subsets of agents are picked, and the group size $k$. In general, all results except Theorem~\ref{thm:inbetween} hold for any variant of Alg.~\ref{alg:dec} as long as each good is guaranteed to eventually get to one of the agents in the argmax on the right-hand side of Eq.~\eqref{math:dec property}.

Therefore, we can define the decentralized process in terms of the property this equilibrium should satisfy, and think of decentralized allocations as analogous to states of equilibrium, without getting bogged down by implementation details. In particular, we define the \emph{decentralized solution} as any allocation $\DEC$ satisfying the property from Equation~\eqref{math:dec property}.

\section{Social Welfare Trade-offs}\label{sec:SW tradeoffs}
In this section we characterize scenarios where the centralized and decentralized solutions succeed in producing allocations with high Nash welfare. Broadly, we will show how both mechanisms' approximation ratios of the maximum Nash welfare change depending on the ratio between  valuations for $G$ and endowments. As endowments grow relative to $G$, the decentralized solution performs better (Theorem~\ref{thm:MA approx MNW}), and as they shrink relative to valuations for individual goods, the centralized solution performs better (Theorem~\ref{thm:CPLI approx MNW}). Perhaps most strikingly, when endowment/valuation ratios are roughly between these extremes, we find a class of examples where a hybrid mechanism that is ``in between'' the centralized and decentralized solutions outperforms both (Theorem~\ref{thm:inbetween}).

As motivation and intuition for these results, we first note that both solutions can approximate the maximum Nash welfare arbitrarily poorly, as we show in the following two examples.

\begin{example}[The decentralized solution can approximate the maximum Nash welfare arbitrarily poorly] 
Suppose we have a fair division instance with $n$ agents where $e_1 = 1$ and $e_i = 1 + \epsilon$ for all $i>1$, where $\epsilon$ is a small constant. Suppose we have $n$ identical goods that are each valued at $1$ by all agents --- i.e., $v_j(g_k) = 1$ for all $j \in N$, $k \in G$. Then the decentralized solution, which only makes allocative decisions on a per-good basis, will assign all of the goods to agent $1$, yielding a Nash welfare of 
\[\log(n+1) + (n-1) \cdot \log(1+\epsilon) \approx \log(n+1) \,.\]
However, the maximum Nash welfare is around $n\log(2)\allowbreak = n$, which is achieved when each agent gets one good. The centralized solution gives this allocation. So the decentralized solution approximates the maximum Nash welfare at a ratio of around
$\frac{\log(n+1)}{n}$,
which goes to 0 as $n$ grows.
\end{example}

In the above example, as the value of $G$ grows relative to the value of the endowments, the decentralized solution performs worse. 
Symmetrically, in the below example there is an agent with a large endowment relative to their value for goods and the centralized solution performs poorly.  

\begin{example}[The centralized solution can approximate the maximum Nash welfare arbitrarily poorly]
Consider a two-agent, two-good fair division instance as follows:
\begin{align*}
    &e_1 = 1 &&v_1(g_1) = 1 &&v_1(g_2) = 1/x -\epsilon \\
    &e_2 = 1 &&v_2(g_1) = x &&v_2(g_2) = 1
\end{align*}
where $\epsilon$ is small. Here, the centralized solution will set $\CEN_1 = \set{g_1}$ and $\CEN_2 = \set{g_2}$, yielding a Nash welfare of $2$. The maximum Nash welfare is $\approx \log(x+1/x+2)$, when agent $2$ gets $g_1$ and agent 1 gets $g_2$ (the decentralized solution produces this allocation). The centralized solution's approximation ratio is $\frac{2}{\log(x+1/x+2)}$,
which goes to 0 as $x$ grows.
\end{example}

The above example takes advantage of the fact that the centralized solution optimizes for Nash welfare as if all endowments are 0. However, this modeling choice is not important for the broad contour of our results: in Appendix~\ref{apdx:CEN bad unif e}, we show that for any positive constant $c$, a centralized solution assuming all endowments are $c$ can also approximate the maximum Nash welfare arbitrarily poorly. Additionally, in Appendix~\ref{apdx:CEN bad -two} we show that the centralized solution can perform arbitrarily poorly even if it knows all but two of the endowments, and even if it can freely make assumptions about the values of the two missing endowments. 

Since the centralized solution lacks access to agents' endowments, it may seem intuitive that it ought to perform poorly when disparities in endowments are large. Notably, however, the above example shows that the centralized solution performs poorly even in the absence of such disparities. 

\subsection{Welfare Guarantees for the Decentralized Solution}

Each solution also performs well in its own unique class of scenarios. First, we will show that as endowments grow, the decentralized solution provides an increasingly tight approximation ratio of the maximum Nash welfare. 

To see why, first, note that due to the equivalence noted in Equation~\ref{math:dec equivalence} and proved in Proposition~\ref{prop:decentralized alloc rule}, $\DEC$ maximizes
\[
\sum_{i \in N} \sum_{g \in \DEC_i} \log(v_i(g) + e_i) - \log(e_i) \,.
\]
For $A_i \subseteq G$, let $u_i(A_i) = \sum_{g \in A_i} \log(v_i(g) + e_i) - \log(e_i)$. Using this new notation, we have that $\DEC$ maximizes $\sum_{i \in N} u_i(\DEC_i)$. Note that $u$ is similar to an alternative function $u^*$:
\[u^*_i(A_i) = \log(v_i(A_i) + e_i) - \log(e_i)\]
Maximizing $\sum_i u^*_i(A_j)$ would maximize Nash welfare, since $\log(e_i)$ is merely a constant. Therefore, in order to show how well the decentralized solution approximates the maximum Nash welfare, we will first show how well $u_i(A_i)$ approximates $u^*_i(A_i)$.

Intuitively, the relative error ought to at least partially depend on the magnitude of $e_i$, since log is concave: as $e_i$ grows, there are diminishing marginal returns from adding in extra value on top. Our next two results quantify this intuition.\footnote{Omitted proofs can be found in the appendix.}

\begin{proposition}\label{prop:u geq u star}
For all agents $i$, $u_i(A_i) \geq u^*_i(A_i)$.
\end{proposition}

\begin{proposition}\label{prop:u approx u star}
For agent $i$, let $C_i = v_i(A_i)/e_i$. Then
\[
\frac{u_i(A_i)}{u^*_i(A_i)}
\leq
\frac{C_i + \frac{C_i^3}{3}}{\ln(1 + C_i)} \,.
\]
\end{proposition}

This result implies that, e.g., if $C_i \leq 0.1$, we have at most 5.271\% relative error of $u(A_i)$ as an approximation of $u^*_i(A_i)$. For $C_i < 1.129$, we also have the looser but more interpretable bound
\[\frac{u_i(A_i) - u^*_i(A_i)}{u^*_i(A_i)} < C_i \,.\]


Propositions~\ref{prop:u geq u star}~and~\ref{prop:u approx u star} pave the way for our main result about the ability of $\DEC$ to approximate the maximum Nash welfare.
For simplicity, for any allocation $A$ let $u^*(A) = \sum_{i \in N} u^*_i(A_i)$ and let $u(A) = \sum_{i \in N} u_i(A_i)$. Furthermore, let $f\colon \Reals_{>0} \to (1,\infty)$ be defined by $f(x) = \frac{x + x^3/3}{\ln(1 + x)}$ and let $z = f\Bigl(\max_i\frac{v_i(G)}{e_i}\Bigr)$.

\begin{theorem}\label{thm:MA approx MNW}
    Assume $\sum_i\log(e_i) \geq 0$. Then
    \[z \cdot \NW\bigl(\DEC\bigr) \geq \NW\bigl(\OPT\bigr) \,.\]
\end{theorem}
\begin{proof}        
First, note that for $i \in N$,
\[
u_i\bigl(\DEC_i\bigr) \leq u^*_i\bigl(\DEC_i\bigr) \cdot f \biggl(\frac{v_i(A_i)}{e_i}\biggr)
\leq u^*_i\bigl(\DEC_i\bigr) \cdot f \biggl(\frac{v_i(G)}{e_i}\biggr) \\
\leq u^*_i\bigl(\DEC_i\bigr) \cdot z 
\]
where the first inequality is due to Proposition~\ref{prop:u approx u star}, the second inequality is due to $f$ being non-decreasing, and the third inequality is due to the definition of $z$.
Summing over all agents yields
\begin{equation}\label{math:MA approx1}
    u\bigl(\DEC\bigr) \leq u^*\bigl(\DEC\bigr) \cdot z \,.
\end{equation}
Additionally, we have
\begin{equation}\label{math:MA approx2}
u^*\bigl(\OPT\bigr) \leq u\bigl(\OPT\bigr)
\end{equation}
by Proposition~\ref{prop:u geq u star}.
By the definition of the decentralized solution, $\DEC \in \argmax_{A \in \textbf{A}} u(A)$. In particular, $u(\OPT) \leq u(\DEC)$. Combining this with Eqs.~\ref{math:MA approx1}~and~\ref{math:MA approx2}, we get
\[u^*\bigl(\OPT\bigr) \leq u^*\bigl(\DEC\bigr) \cdot z \,.\]

\noindent Letting $k = \sum_i\log(e_i)$, we have
\[
u^*\bigl(\OPT\bigr) + k \leq z \cdot u^*\bigl(\DEC\bigr) + k 
\leq z \cdot u^*\bigl(\DEC\bigr) + z\cdot k 
= z \cdot \bigl(u^*\bigl(\DEC\bigr) + k\bigr)
\]
where the second inequality follows from the assumption that $k \geq 0$ and the fact that $z > 1$.
Finally, we have
\[\sum_i \log\bigl(v_i\bigl(\OPT_i\bigr) + e_i\bigr) \leq z \cdot \sum_i \log\bigl(v_i\bigl(\DEC_i\bigr) + e_i\bigr)\]
by the definition of $u^*$, yielding
    \[z \cdot \NW\bigl(\DEC\bigr) \geq \NW\bigl(\OPT\bigr)\]
by the definition of Nash welfare.\qed
\end{proof}

Of course, the guarantee provided by this theorem depends on $z$, which itself depends on $\max_i v_i(G) / e_i$. As $\max_i v_i(G) / e_i$ approaches 0, $z$ approaches 1 and $\NW(\DEC)$ approaches $\NW(\OPT)$.
However, we get interesting bounds before then, as shown in Table~\ref{table:MA approx bounds}.

\begin{table}[bhpt]
\centering
\begin{tabular}{ cccccc } 
\toprule
$\max_i v_i(G)/e_i$ & $1.0$ & $0.8$ & $0.6$ & $0.4$ & $0.2$ \\
\midrule
$z$ & $1.92$ & $1.65$ & $1.43$ & $1.25$ & $1.11$ \\
Approx. ratio of MNW &  $0.52$ & $0.6$ & $0.7$ & $0.8$ & $0.9$\\
\bottomrule
\end{tabular}
\caption{Approximate values of $z$ and corresponding approximation ratios for different values of $\max_i v_i(G)/e_i$.}
\label{table:MA approx bounds}
\end{table}

For example, the first column of Table~\ref{table:MA approx bounds} tells us that when one agent values their endowment and $G$ equally and all other agents weakly prefer their endowments to $G$,
$\DEC$ provides better than a 1/2-approximation of the maximum Nash welfare. When each agent values their endowment at least 5 times as much as $G$, $\DEC$ provides a $0.9$-approximation of the maximum Nash welfare. If we think of endowments are as stand-ins for general levels of socioeconomic status, such values for $z$ start to feel quite reasonable in many real-world scenarios. 

One might wonder if Theorem~\ref{thm:MA approx MNW} could apply to the centralized solution as well. After all, if endowments are very large compared to valuations, then an agent's endowment will make up most of their utility no matter what goods they receive -- so maybe one is always guaranteed a good approximation ratio. 

While this is true in the limit as endowments approach infinity (and $\max_i v_i(G)/e_i \to 0$), we will show a strict gap between the worst-case performances of $\DEC$ and $\CEN$ for values of $\max_i v_i(G)/e_i$ as small as 0.194. In particular, this means that the centralized solution cannot achieve any of the approximation guarantees described in Table~\ref{table:MA approx bounds}.

\begin{theorem}\label{thm:CEN counterexamples}  
    For every $x \geq 0.194$, there is an infinite family of fair division instances with $\max_i v_i(G)/e_i = x$ and $z \cdot \NW\bigl(\CEN\bigr) < \NW\bigl(\OPT\bigr)$.
\end{theorem}

\subsection{Welfare Guarantees for the Centralized Solution}

We now show approximation bounds for $\CEN$. Mirroring Theorem~\ref{thm:MA approx MNW}, which shows that $\DEC$ performs well when endowments are high relative to valuations for $G$, the following theorem shows that $\CEN$ performs well when endowments are low relative to valuations for individual goods. Let $\alpha$ denote the maximum ratio between any agent's endowment and their valuation for a good; i.e.,  $\alpha = \max_{i \in N, g \in G, v_i(g) > 0} \frac{e_i}{v_i(g)}$.

\begin{theorem}\label{thm:CPLI approx MNW}
     Suppose $\CEN$ and $\OPT$ both assign at least one good to every agent, and every good is valued positively by at least one agent. Then
    $\NW\bigl(\CEN\bigr) > \NW\bigl(\OPT\bigr) - n \cdot \log(1+\alpha)$.
\end{theorem}

Just as Theorem~\ref{thm:MA approx MNW}'s approximation bound doesn't apply to $\CEN$, Theorem~\ref{thm:CPLI approx MNW}'s bound doesn't apply to $\DEC$. To see why, consider what happens when all endowments are equal and small. As they go to zero, Theorem~\ref{thm:CPLI approx MNW} guarantees that the centralized solution will maximize Nash welfare. However, whenever all endowments are equal, the decentralized solution will maximize utilitarian social welfare (i.e., maximize $\sum_i v_i(A_i)$). The Nash welfare of the resultant allocation can be an arbitrarily poor approximation of the maximum.



\subsection{A Hybrid Mechanism can Outperform Both}

Finally, we show that the optimal solution may be ``in between'' the centralized and decentralized solutions. Recall that $\DEC$ can be defined as the convergent state of a particular decentralized exchange process. However, if we start with the centralized solution and run a \textit{non-convergent} amount of decentralized exchange, then under certain conditions this intermediate allocation has higher Nash welfare than both $\CEN$ and $\DEC$.  While the conditions in the Theorem are restrictive, in Section~\ref{sec:exps} we empirically find this occurs more broadly.

\begin{theorem}\label{thm:inbetween}
    Consider a fair division instance where $n = m$ and all agents have value $1$ for all goods.
    Suppose endowments are drawn from the uniform distribution over $(0, M]$.
    Then, via exchange among dyads according to the decentralized process\footnote{I.e., running Algorithm~\ref{alg:dec} with $k=2$}, starting from an initial allocation where each agent holds one good, Nash welfare increases in the first step of the decentralized process and decreases in the last step with probability at least
    \begin{align*}
            \biggl(1 - \biggl(1 - \frac{1}{n}\biggr)^n\biggr) \cdot \frac{(n-2)\Bigl(M - \frac{M}{n} - 1\Bigr)^2}{n\Bigl(M - \frac{M}{n}\Bigr)^2} \cdot \min\biggl(\biggl(\frac{n - 1}{M}\biggr)^n, 1\biggr)\;.
    \end{align*}%
\end{theorem}

\begin{corollary}
If\/ $M \in [18, n-1]$, then Nash welfare increases in the first step of the decentralized process and decreases in the last step with probability at least $\frac{1}{2}$.
\end{corollary}

\section{Fairness Trade-offs}\label{sec:fairness tradeoffs}
It is well known that the centralized solution is guaranteed to be \EFo{}. However, this notion of fairness ignores endowments and provides only weak guarantees when, for example, a small number of goods are generally more desirable than others.  As an alternative, we have introduced the notion of \EREF{}, which requires, for example, that those desirable goods be allocated among those with the lowest endowments. As we will explain, the decentralized solution cannot achieve \EREF{} in every case.  However, we identify cases where it is guaranteed to hold and show that it holds in expectation (also in some cases with high probability) when valuations are drawn i.i.d.\ for each good.

Before proving fairness guarantees of the decentralized solution, we observe that depending on what the endowments look like, the decentralized solution can embody a spectrum of social welfare objectives, from strictly utilitarian to strictly egalitarian. When all endowments are equal, the decentralized solution simply maximizes utilitarian social welfare. On the other hand, with high disparity in endowments, the decentralized solution will behave in a more egalitarian way by funneling all of the goods to the agents with low endowments. In some circumstances, such behavior even maximizes egalitarian social welfare (i.e., maximizes the utility of the worst-off agent).

When the disparity in endowments is in between these extremes, the decentralized solution naturally trades off between utilitarian and egalitarian objectives in the same way as other Nash welfare-based approaches. In particular, this means that in some cases it will not satisfy \EREF{}. Consider two agents $i$ and $j$, and a good $g$: if $e_i$ is slightly smaller than $e_j$ but $v_i(g)$ is much larger than $v_j(g)$ then $g$ will be allocated to $i$ despite the lower endowment, which can lead to a violation of \EREF{}.  Thus, we turn to results characterizing conditions under which it can be guaranteed, either unconditionally or probabilistically.

First, we provide conditions under which \EREF{} can be unconditionally guaranteed. Our first result focuses on valuations that are in the set $\set{0} \cup [\ell, h]$ for $0 < \ell \leq h$. In other words, our theorem applies when every nonzero valuation is between some lower bound $\ell$ and upper bound $h$. We refer to such valuations as $0$--$\ell$--$h$ valuations. 

\begin{theorem}[Partial \EREF{} with $0$--$\ell$--$h$ valuations]\label{thm:1wayEF deterministic}
If valuations are $0$--$\ell$--$h$ then 
\[
e_i < \frac{\ell}{h} e_j \implies v_i\bigl(\DEC_i\bigr) \geq v_i\bigl(\DEC_j\bigr) \,.
\]
\end{theorem}
\begin{proof}
Suppose that $v_i(g) \in \set{0} \cup [\ell, h]$ for all $i \in N$ and $g \in G$, with $\ell \leq h$. Then if $e_i < \frac{\ell}{h} \cdot e_j$, then $\DEC_j$ cannot contain any good $g$ that agent $i$ has positive value for. This is because $v_i(g) \in [\ell, h]$ implies 
\[
\frac{v_i(g)}{e_i} \geq \frac{\ell}{e_i} > \frac{h}{e_j} \geq \frac{v_j(g)}{e_j} 
\]
so $g \in \DEC_j$ would contradict the definition of $\DEC$. So $v_i(A_j) = 0$. \EREF{} follows from observing that $v_i(A_i) \geq 0$.\qed
\end{proof}

This theorem exposes a trade-off between the expressiveness of valuations and the strength of the \EREF{} guarantee provided by $\DEC$. In particular, if $\ell = h$ (i.e., valuations are drawn from the two-element set $\set{0,h}$) then $\DEC$ satisfies \EREF{} unconditionally. However, as $\ell$ and $h$ grow farther apart, \EREF{} may not apply for some pairs of agents.

Our next theorem shows that $\DEC$ is \EREF{} in expectation when valuations for each good are drawn i.i.d. from a distribution specific to that good.

\begin{theorem}[\EREF{} in expectation]\label{thm:1wayEF expectation}
If valuations for each good $g$ are drawn i.i.d. from a distribution $\mathcal{D}_g$ then
\[
e_i \leq e_j \implies
\E\bigl[v_i\bigl(\DEC_i\bigr)\bigr] \geq \E\bigl[v_i\bigl(\DEC_j\bigr)\bigr]
\]
where the expectation is over the randomness in the valuations.
\end{theorem}
The proof of this theorem requires the following technical Lemma.
\begin{lemma}\label{lemma:expected vals}
Let $X$ and $Y$ be independent random variables and $Z$ any random variable. Assume that $f_Y$, the probability density function of $Y$, exists. Then
\[
\E[X \mid Z \land (X \geq Y)] \geq \E[X \mid Z \land (X \leq Y)] \,.
\]
\end{lemma}

Before proving Theorem~\ref{thm:1wayEF expectation}, we'll discuss one way of describing allocations and valuations for allocated bundles as random variables. We will assume that for $i \in N$ and $g \in G$, there exists a random variable $v_i(g)$ denoting $i$'s value for $g$. Let $\ind(g \in \DEC_i)$ denote the indicator variable taking a value of $1$ if the decentralized solution places $g$ in $
\DEC_i$, and $0$ otherwise. Then, for all $i,j \in N$, we can let $v_i(\DEC_j)$ be a random variable denoting $i$'s value for $j$'s bundle under the decentralized solution. But notice that we can also write
\[v_i(\DEC_j) = \sum_{g \in G} v_i(g) \cdot \ind(g \in \DEC_j) \,.\]
We will use this representation in the proof.

\begin{proof}[Theorem~\ref{thm:1wayEF expectation}]
For notational simplicity, let $\DEC$ be denoted simply as $A$.
First, we'll show that we can focus on the case of one good. We have
\begin{align*}
&\E[v_i(A_i) - v_i(A_j)] 
= \sum_{g \in G} \bigl( \E[v_i(g) \cdot \ind(g \in A_i)] - \E[v_i(g) \cdot \ind(g \in A_j)] \bigr)
\end{align*}
by linearity of expectation, so to prove the theorem, it suffices to show
\[\E[v_i(g) \cdot \ind(g \in A_i)] \geq \E[v_i(g) \cdot \ind(g \in A_j)]\]
for an arbitrary good $g$.

Next, we'll massage these expectations a bit. We have the following:
\begin{align*}
    \displaystyle
    &\E[v_i(g) \cdot \ind (g \in A_i)] \\
    ={} &\E[v_i(g) \cdot \ind (g\in A_i) \mid g \notin A_i] \cdot \Pr(g \notin A_i) \\
    &\phanrel + \E[v_i(g) \cdot \ind (g \in A_i) \mid g \in A_i] \cdot \Pr(g \in A_i) \\
    ={} &\E[0 \mid g \notin A_i] \cdot \Pr(g \notin A_i) 
    + \E[v_i(g) \mid g \in A_i] \cdot \Pr(g \in A_i) \\
    ={} &\E[v_i(g) \mid g \in A_i] \cdot \Pr(g \in A_i)
\end{align*}
Likewise, we also have $\E[v_i(g) \cdot \ind (g \in A_j)] = \E[v_i(g) \mid g \in A_j] \cdot \Pr(g \in A_j)$.
%
Thus our proof will proceed by showing first that $\Pr(g \in A_i) \geq \Pr(g \in A_j)$ and second that $\E[v_i(g) \mid g \in A_i] \geq \E[v_i(g) \mid g \in A_j]$.

\bn \textit{1. Showing that $\Pr(g \in A_i) \geq \Pr(g \in A_j)$.}

\mn By the definition of the decentralized mechanism, $g$ is put into $A_i$ if and only if 
\begin{align}\label{math:alloc rule}
\frac{v_i(g)}{e_i} \geq \frac{v_k(g)}{e_k} 
\end{align}
for all agents $k \ne i$. Let $F$ be the CDF of the distribution $\mathcal{D}_g$ that valuations for good $g$ are drawn from. Then
\begin{align*}
\Pr(g \in A_i \mid v_i(g) = x) &= \Pr\biggl(\bigwedge_{k \neq i} v_k(g) \leq x \frac{e_k}{e_i} \biggr) =\prod_{k\neq i} F\biggl(x\frac{e_k}{e_i}\biggr)
\end{align*}
where the first equality follows from the definition of the decentralized mechanism and from rearranging Eq.~\ref{math:alloc rule}, and the second equality follows from the independence of the $v_k(g)$ terms. Likewise,
\[\Pr(g \in A_j \mid v_j(g) = x) = \prod_{k \neq j} F\biggl(x\frac{e_k}{e_j}\biggr) \,.\]
Note that for nonnegative $x$ and all agents $k$, we have
\[e_i \leq e_j \implies x \frac{e_k}{e_i} \geq x \frac{e_k}{e_j} \implies F\biggl(x \frac{e_k}{e_i}\biggr) \geq F\biggl(x \frac{e_k}{e_j}\biggr) \]
where the second implication follows from the monotonicity of CDFs. We also have
\[e_i \leq e_j \implies x\frac{e_j}{e_i} \geq x\frac{e_i}{e_j} \implies F\biggl(x\frac{e_j}{e_i}\biggr) \geq F\biggl(x\frac{e_i}{e_j}\biggr) \,.\]
Therefore, 
\begin{align*}
&\phanrel \Pr(g \in A_i \mid v_i(g) = x) = \prod_{k \neq i} F\biggl(x\frac{e_k}{e_i}\biggr)
= F\biggl(x\frac{e_j}{e_i}\biggr) \cdot \prod_{k \neq i,j} F\biggl(x\frac{e_k}{e_i}\biggr) \\
&\geq F\biggl(x\frac{e_j}{e_i}\biggr) \cdot \prod_{k \neq i,j} F\biggl(x\frac{e_k}{e_j}\biggr) 
\geq F\biggl(x\frac{e_i}{e_j}\biggr) \cdot \prod_{k \neq i,j} F\biggl(x\frac{e_k}{e_j}\biggr) 
=\prod_{k \neq j} F\biggl(x\frac{e_k}{e_j}\biggr)\\
&=\Pr(g \in A_j \mid v_j(g) = x) \,.
\end{align*}
Let $f$ be the PDF of $\mathcal{D}_g$.
Via the law of total probability, we can write $\Pr(g \in A_i)$ as an integral:
\begin{align}\label{math:int Pr g in A_i}
    \Pr(g\in A_i) 
    = \int_0^{\infty} \Pr(g \in A_i \mid v_i(g) = x) f(x) \, dx
\end{align}
Likewise:
\begin{align}\label{math:int Pr g in A_j}
    \Pr(g\in A_j) 
    = \int_0^{\infty} \Pr(g \in A_j \mid v_j(g) = x) f(x) \, dx 
\end{align}
Finally, the integrand in Eq.~\ref{math:int Pr g in A_i} is always greater than or equal to the integrand in Eq.~\ref{math:int Pr g in A_j}, so we have $\Pr(g\in A_i) \geq \Pr(g \in A_j)$ as desired.

\bn \textit{2. Showing that $\E[v_i(g) \mid g \in A_i] \geq \E[v_i(g) \mid g \in A_j]$.} 

\mn We first reinterpret the allocation found by the decentralized mechanism as follows. Suppose we draw a valuation for $g$ for each agent $i$, then divide the valuations by the associated endowments of each agent, and then finally rank the agents in order of these scaled valuations:
\[
\frac{v_{(1)}(g)}{e_{(1)}} \leq \frac{v_{(2)}(g)}{e_{(2)}} \leq \cdots \leq \frac{v_{(n)}(g)}{e_{(n)}} \,.
\]
Then the decentralized mechanism simply assigns the good to agent $(n)$.
That means we can rewrite one of our expectations as follows:
\[
\E[v_i(g) \mid g \in A_j] = \E[v_i(g) \mid j = (n)] \,.
\]
We now also have
\begin{align*}
    &\phanrel \E[v_i(g) \mid j = (n)] \\
    &\leq \E[v_i(g) \mid j = (n)\ \text{and}\ i = (n-1)] \\
    &
    = \E\biggl[v_i(g) \mathrel{\biggm|}   v_i(g) \geq \frac{v_k(g) e_i}{e_k}\ \forall k \neq j 
    \text{and}\ v_i(g) \leq \frac{v_j(g) e_i}{e_j} \biggr] 
\end{align*}
where the inequality follows from the fact that conditioning on $i$ having a higher ranking can only increase $v_i$, and the equality follows from the definition of $j = (n)$ and $i = (n-1)$.
Furthermore,
\begin{align*}
    \E[v_i(g) \mid g \in A_i] &= \E[v_i(g) \mid i = (n)] 
    = \E\biggl[v_i(g) \mathrel{\biggm|} v_i(g)\geq \frac{v_k(g) e_i}{e_k}\ \forall k \biggr] \,.
\end{align*}
Let $X = v_i(g)$ and $Y = \frac{v_j(g) e_i}{e_j}$, and let $Z$ be the event that $v_i(g)\geq \frac{v_k(g) e_i}{e_k}$ for all $k \ne j$. Then, we have the following.
\begin{align*}
    \E[v_i(g) \mid g \in A_i] &= \E[X \mid Z \land (X \geq Y)] \\
    \E[v_i(g) \mid g \in A_j] &\leq \E[X \mid Z \land (X \leq Y)]
\end{align*}
Finally, by Lemma~\ref{lemma:expected vals}, since $f_Y$ exists, we have
\[\E[X \mid Z \land (X \geq Y)] \geq \E[X \mid Z \land (X \leq Y)] \]
completing the proof.\qed
\end{proof}

The decentralized solution provides a stronger fairness guarantee in expectation when valuations are drawn from uniform distributions.

\begin{corollary}\label{cor:EREF uniform}
    Suppose $\mathcal{D}_g$ is a uniform distribution $\mathcal{U}_{[0, M_g]}$ for all $g$. Then $e_i \leq e_j$ implies
    \[\E\bigl[v_i\bigl(\DEC_i\bigr)\bigr] \geq \frac{e_j}{e_i} \cdot \E\bigl[v_i\bigl(\DEC_j\bigr)\bigr] \,.\]
\end{corollary}

Lastly, we demonstrate that \EREF{} holds with high probability when valuations for each good are drawn from a (possibly distinct) uniform distribution.
\begin{theorem}[\EREF{} with high probability]\label{thm:1wayEF probability}
Suppose valuations for each good $g$ are drawn independently from a uniform distribution $\mathcal{U}_{[0, M_g]}$. Then for any two agents $i$ and $j$ with $e_i \leq e_j$,
\[\Pr\Bigl(v_i\bigl(\DEC_i\bigr) \geq v_i\bigl(\DEC_j\bigr)\Bigr) \geq \Biggl(1 - \frac{\prod e_k}{n e_j^n}\Biggr)^m \]
where the probability is over randomness in valuations.
\end{theorem}

\section{Simulations}\label{sec:exps}
Here, we analyze randomized fair division instances to understand how our mechanisms behave in the presence of disparities in endowments. We also study two additional mechanisms that represent partial and combined versions of the centralized and decentralized approaches. In particular, we study what happens when we only run Algorithm~\ref{alg:dec} for three rounds, with $k = n/3$ randomly chosen agents exchanging goods in each round. We simulate this process starting from two different ``baseline'' allocations: $\CEN$, and a random allocation. The former of these additional mechanisms is meant to capture the way that centralized and decentralized systems often intermingle in practice, and the latter is chosen to test what happens when we add more randomness and uncertainty into the decentralized process.

For a given fair division instance, let $\OPT$ represent an allocation that maximizes Nash welfare and let $\NUL$ represent the ``empty'' allocation, where no goods are given to any agent. In this section we study the ratio
\begin{align}\label{math:exp ratio}
\frac{\NW(A) - \NW(\NUL)}{\NW(\OPT) - \NW(\NUL)}
\end{align}
for various choices of $A$. In other words, we measure the approximation ratios of these allocative schemes in terms of the \emph{marginal} increase in welfare they provide over the endowments. 
In our simulations we generated random fair division instances with endowments drawn from a uniform distribution over integers between $1$ and $e_{\max}$ (a parameter we varied over the course of each experiment). 
Valuations were drawn uniformly from the set of integral valuations summing to 500. For each of several choices of $m$ and $n$, we ran 100 trials each at various values of $e_{\max}$.

\begin{figure}[bht]  
    \centering
    \includegraphics[width=0.49\linewidth]{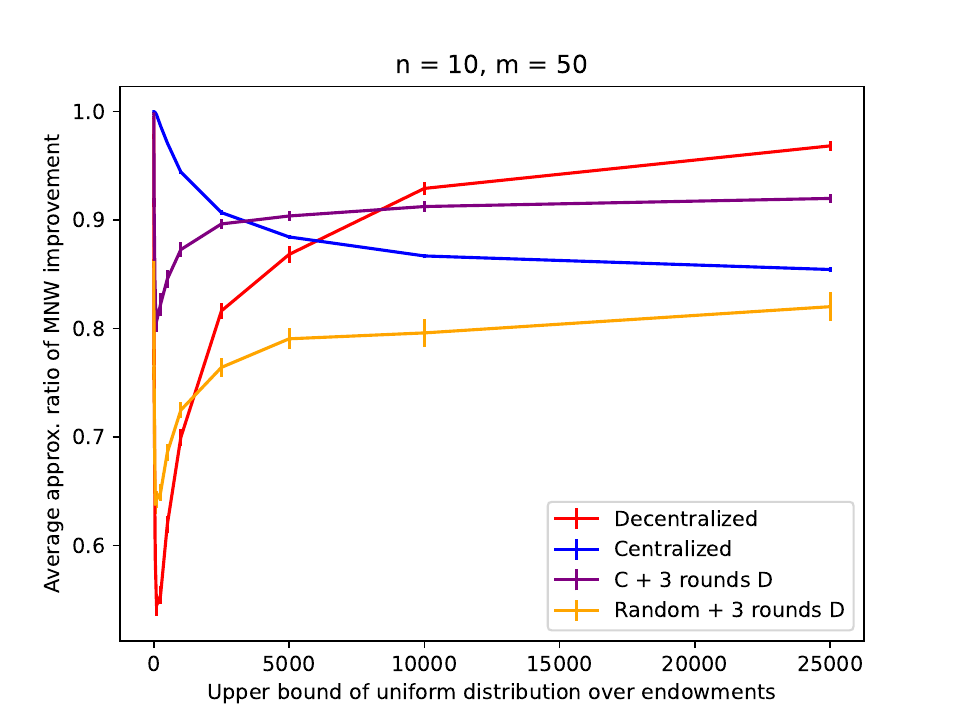}
    \includegraphics[width=0.49\linewidth]{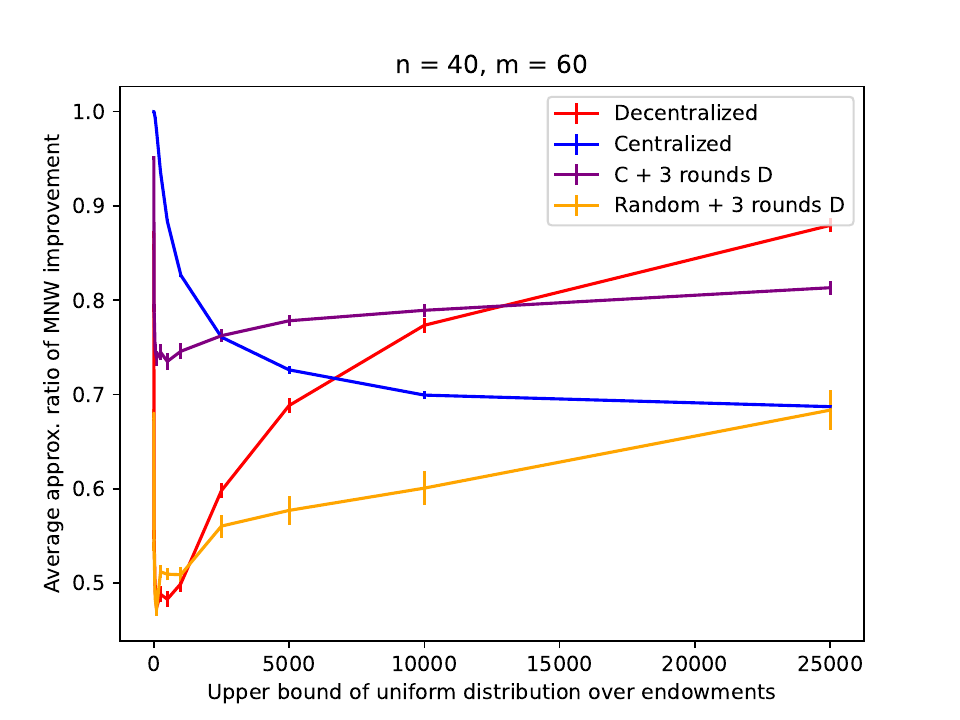}
    \caption{Approximation ratios of the allocative schemes as a function of disparity in endowments. Tick marks show standard error.}
    \label{exp:disparity1}
\end{figure}

For each trial, we calculated ratio~\eqref{math:exp ratio} for the above four mechanisms.
Figure~\ref{exp:disparity1} shows their approximation ratios as a function of disparities in endowments (i.e. $e_{\max}$). Additional plots reflecting other problem sizes are in the appendix.

These data reveal that $\DEC$ tends to perform better when there is more disparity in endowments, overtaking $\CEN$ when $e_{\max} \approx 5000$ in all cases.  We contend that in the real world, it is not unusual to see endowments this large, or larger, compared to the value of goods being redistributed.

These data also show that starting from $\CEN$ and then running even a small number of rounds of decentralized exchange can have a meaningful impact on Nash welfare, raising the approximation ratio by around 10 percentage points in the best case, and only negatively impacting the welfare when endowment disparity is relatively small. Following Theorem~\ref{thm:inbetween}, all of our simulations feature a portion of the $x$ axis in which this hybrid approach outperforms both $\CEN$ and $\DEC$. This may point to a useful synergy of decentralized and centralized techniques in practice. 

\section{Conclusion}
This paper develops a stylized model capturing important aspects of decentralized networks and contrasting it with a model of a centralized allocation. While each model has strengths and drawbacks, our results suggest that decentralized protocols like ours are not merely niche techniques but applicable to a broad range of scenarios where the values of the goods to be distributed are neither so small relative to endowments that the problem becomes unimportant nor so large relative to endowments that the values of the endowments become unimportant, making the centralized solution better. 

This work was inspired by networks where principles of cooperation and mutual trust have allowed communities to accomplish interesting and useful allocative goals which might not be possible in traditional economic frameworks. These principles both 1) constitute an important re-allocative force in practice, and 2) provide a powerful normative framework through which fairness research can be understood and improved. Analyzing fair division through this framework -- in which humans are inherently social and connected -- is essential for understanding of how mechanisms of all types can succeed.

\bibliographystyle{splncs04}
\bibliography{references}

\appendix


\section{In-depth Explanation of the Decentralized Allocation Mechanism}\label{apdx:dec model explanation}
Here we give a full explanation of our decentralized allocation mechanism. While this mechanism is not meant to describe any specific real-world phenomenon, it is meant to capture the key features of decentralized networks observed in Section~\ref{sec:intro}: namely, that individuals in these networks 1) can forsake utility maximization in the pursuit of other goals, 2) can have access to additional knowledge, and 3) often use simplified decision making strategies, rather than computationally intensive ones. We also rely on the observation that in allocative tasks, individuals prefer social welfare-increasing changes over changes that only reduce inequality~\cite{charness2002understanding}. 

Our decentralized mechanism proceeds in a series of time steps. Given a fair division instance, the agents start with some allocation $A^0$, and at each time step $t$, some subset $S^t \subseteq N$ of the agents redistributes the resources they currently hold among themselves. That redistribution induces a new allocation $A^{t+1}$, and we proceed to the next time step. At each time step, goods are distributed according to an approximation of Nash welfare maximization. In particular, when the subset $S^t$ interacts, each good $g$ held by some member of $S^t$ is given to an agent in $\argmax_{i \in S^t} \frac{v_i(g)}{e_i}$, with ties broken randomly. As shown in Proposition~\ref{prop:decentralized alloc rule}, this is equivalent to giving $g$ to a member of $\argmax_{i \in S^t} \log(v_i(g) + e_i) - \log(e_i)$. As in Section~\ref{sec:SW tradeoffs}, we define $u_i(g) := \log(v_i(g) + e_i) - \log(e_i)$ for single goods $g$ and $u_i(X) = \sum_{g\in x} \log(v_i(g) + e_i) - \log(e_i)$ for bundles $X \subseteq G$.

An agent in $S^t$ who starts with a good $g$ need not give it away, if they happen to be an agent $i$ maximizing $u_i(g)$ within that local subset. However, consistent with the evidence from behavioral economics and the other literature mentioned in Section~\ref{sec:intro}, an agent may also leave the interaction with a less valuable bundle than they entered with. Moreover, agents can ``see'' endowments when calculating $u_i(g)$ for a given $i$ and $g$, consistent with the observation that agents have access to more local information than centralized entities. For any good $g$, finding an agent $i\in S^t$ maximizing $u_i(g)$ takes time linear in $|S^t|$, capturing the observation from the literature discussed in Section~\ref{sec:intro} that individuals in local contexts can be resource constrained.

Formally, we model one round of the decentralized allocation process as a randomized algorithm that, starting with some initial allocation, chooses $k$ agents and redistributes goods amongst them.  We fix $k \in \set{2, \dots, n}$ for this discussion and denote by $S_k$ the collection of subsets of $N$ with size $k$.  Let $A^0$ be the initial allocation. At each step $t\ge 0$, with allocation $A^t$, a set $S^t \in S_k$ is chosen uniformly at random to re-allocate their goods and create a new allocation $A^{t+1}$.



Given $A^t$ and $S^t$, the set of \emph{locally feasible} re-allocations that occur only among $S^t$'s members is defined as the set of allocations where all of non-$S^t$ members have the same allocation as they do in $A^t$ --- thereby capturing the idea that only members of $S^t$ should be exchanging goods at time step $t$. We denote this set of locally feasible allocations by $L(A^t,S^t)$:
\[
L(A^t, S^t) = \set{A \in \mathbf{A} \suchthat i \notin S^t \implies A_i = A^t_i} \,.
\]

Next, given an allocation $A^t$ and a set $S^t$, the redistribution reached by the members of $S^t$ is an allocation, denoted by $R(A^t, S^t)$, that maximizes the sum of the $u_i$ terms among the locally feasible allocations:
\[
R(A^t, S^t) \in \argmax_{A \in L(A^t, S^t)} \sum_{i \in S^t} u_i(A_i) \,.
\]
If the set given by the $\argmax$ is not a singleton, in our simulations we randomly choose one element of it.

Since each local re-allocation finds an allocation $A^{t+1}$ maximizing $\sum_i u_i(A^{t+1}_i)$ (given the local feasibility constraints defined by $L(A^t, S^t)$), as $t$ grows the decentralized process should converge to an allocation where every good $g$ goes to the agent who has the globally highest $u_i(g)$ for it.
Equivalently, this is an allocation that maximizes $\sum_i u_i(A_i)$. Furthermore due to Proposition~\ref{prop:decentralized alloc rule} this is also an allocation where $g \in A_i \iff i \in \argmax_{j \in N} \frac{v_j(g)}{e_j}$, matching our definition of the decentralized solution from Section~\ref{subsec:decentralized def}.
The following proposition establishes that as $t$ grows, the probability that the decentralized process maximizes $\sum_i u_i(A_i)$ goes to 1.

\begin{proposition}\label{prop:prob of MA convergence}
For any fair division instance and initial allocation $A^0$,
\[\Pr\Bigl(A^t \in \argmax_{A\in \mathbf{A}} \sum_{i} u_i(A_i)\Bigr) \geq 1 - m \biggl(1 - \frac{k^2 - k}{n^2 - n}\biggr)^t \;. \]
\end{proposition}
\begin{proof}
We'll start by describing the probability that an individual good $g$ reaches an agent in $\argmax_{i \in N} u_i(g)$. This probability is minimized when there is only one agent in this set, since the probability of a union of events is always greater than the probability of one of those events. So suppose there is one member of this set, and call this agent $i^*_g$.

Depending on how $A^0$ is chosen, there may be some probability that $i^*_g$ already starts with $g$. However, if $A^0$ is adversarially chosen then  $i^*_g$ does not start with the good, so we will assume that that is the case. 

At each time step $t' \leq t$, if $i^*$ does not already have the good, there is an $\binom{n-2}{k-2}/\binom{n}{k}$ chance that $i^*_g$ and the current owner of the good are both chosen to be in $S^{t'}$ (recall that $k$ is the size of each set $S^{t'}$). If this event happens, then $i^*_g$ will acquire the good time step $t'$ and then keep the good permanently. If it does not happen, $i^*$ will not acquire the good at that time step. Also note that $\binom{n-2}{k-2}/\binom{n}{k}$ simplifies to $\frac{k^2-k}{n^2-n}$.

Now, define $B_{g}$ as the event that $i^*_g$ does \emph{not} have $g$ after $t$ time steps.
Then
\[\Pr(B_g) = \biggl(1 - \frac{k^2 - k}{n^2 - n}\biggr)^t\]
and
\[\Pr(\cup_{g\in G} B_g) \leq m \biggl(1 - \frac{k^2 - k}{n^2 - n}\biggr)^t\]
follows via a union bound. $\Pr(\cup_{g\in G} B_g)$ is the probability that at least one good does not find its way to the agent who has the highest $u_i$ for it, so $1 - \Pr(\cup_{g\in G} B_g)$ is the probability that all goods end up with the agents with the highest $u_i$ for them. But since $u_i$ is additive for bundles, this condition is also sufficient to guarantee that the resulting allocation maximizes $\sum_{i \in N} u_i(A_i)$.\qed
\end{proof}

Lastly, we show that the two conditions identified in Section~\ref{sec:model} that identify the recipient of a good really are equivalent.

\begin{proposition}\label{prop:decentralized alloc rule}
For every agent $i$ and good $g$,
\[
\argmax_{i\in N} u_i(g) = \argmax_{i \in N} \frac{v_i(g)}{e_i} \,.
\]
where $u_i(g) = \log(v_i(g) + e_i) - \log(e_i)$
\end{proposition}

\begin{proof}
Suppose $i \in \argmax_{i\in N} u_i(g)$. Then for all $j \in N$, we have the following.
\begin{align*}
    &\mathrel{\phantom{\iff}} u_i(g) \geq u_j(g) \\
    &
    \iff \log(v_i(g) + e_i) - \log(e_i)
    \geq \log(v_j(g) + e_j)  - \log(e_j)
     \\
    &\iff \log\biggl(\frac{v_i(g)}{e_i} + 1\biggr) \geq \log\biggl(\frac{v_j(g)}{e_j} + 1\biggr) \\
    &\iff \frac{v_i(g)}{e_i} \geq \frac{v_j(g)}{e_j}  
\end{align*}
where the first transition follows from the definition of $u_i(g)$ and the last follows from the monotonicity of log.\qed
\end{proof}

\section{Proofs and Examples from Section~\ref{sec:SW tradeoffs}}

\subsection{The Centralized Solution Approximates MNW Poorly for Any Assumed Endowment Level}\label{apdx:CEN bad unif e}

Suppose that instead of maximizing $\sum_{i \in N} \log(v_i(A_i))$, the centralized solution instead maximized Nash welfare assuming that all agents had an endowment of $c$ for some constant $c \in \mathbb{R}_{>0}$. Let $A^{\mathrm{CEN},c}$ represent such a centralized solution. In other words, for a given $c$ let

\[A^{\mathrm{CEN},c} 
\in \argmax_{A \in \mathbf{A}}\sum_{i \in N} \log(c + v_i(A_i)) \,.\]

Regardless of the choice of $c$, there exist fair divsion instances where $A^{\mathrm{CEN},c}$ approximates the maximum Nash welfare arbitrarily poorly. Consider an instance with two agents and one good as follows:
\begin{align*}
    &e_1 = 2^x &&v_1(g_1) = 2^x \\
    &e_2 = 1/2^x &&v_2(g_1) = 2^x-\epsilon 
\end{align*}

We will have $A^{\mathrm{CEN},c}_1 = \{g_1\}$ and $A^{\mathrm{CEN},c}_2 = \emptyset$, yielding a Nash welfare of $\log(2)$. Meanwhile, the allocation maximizing Nash welfare gives agent $2$ the good, yielding a Nash welfare of around $\log(1 + 2^{2x})$ ($\DEC$ also gives this allocation). So the approximation ratio achieved by $A^{\mathrm{CEN},c}$ is around $\frac{1}{\log(1 + 2^{2x})} < \frac{1}{2x}$, which goes to 0 as $x$ grows. 

\subsection{The Centralized Solution Approximates MNW Poorly Even With Access to All But Two Endowments}\label{apdx:CEN bad -two}

Here, we show that even if a centralized party knows $n-2$ endowments, it can still generate an allocation that approximates the maximum Nash welfare arbitrarily poorly. 

Without loss of generality, assume that the centralized solution knows $e_3,\dots,e_n$, and let $e_1'$ and $e_2'$ denote the centralized party's assumed values of $e_1$ and $e_2$. We can construct an infinite family of fair division instances where the allocation produced by assuming endowment values $e_1'$ and $e_2'$ is arbitrarily worse than the allocation produced knowing the correct endowment values $e_1$ and $e_2$.

We will require that the good set of this instance $G$ can be partitioned into two sets, $G_1$ and $G_2$ such that
\[
\forall i \notin \{1,2\}, \forall g \in G_1, v_i(g)=0
\]
and
\[
\forall i \in \{1,2\}, \forall g \in G_2, v_i(g)=0 
\]
In other words, none of agents $3$ through $n$ get any positive value from any of the goods in $G_1$, and agents $1$ and $2$ don't get any positive value from any of the goods in $G_2$. This partition allows us to cleanly split the fair division problem into two sub-problems: optimally allocating the goods in $G_1$ to agents 1 and 2, and optimally allocating the goods in $G_2$ to agents 3 through $n$. Let $w$ denote the maximum Nash welfare possible from allocating $G_2$ to agents $3$ through $n$, i.e.
\[
w = \max_{A \in \mathbf{A}}\sum_{i = 3}^{n}\log(v_i(A_i)+e_i) \,.
\]
Without loss of generality, assume that $e_1' \leq e_2'$. Then letting $G_1$ contain a single good $g_1$, we can set $e_1, e_2, v_1(g_1)$ and $v_2(g_2)$ as they are set in the above example in Appendix~\ref{apdx:CEN bad unif e}. The centralized solution will then give $g_1$ to agent 1 and allocate the goods in $G_2$ optimally, yielding a Nash welfare of $w + \log(2)$. Meanwhile, the optimal solution gives $g_1$ to agent 2 and also allocates $G_2$ optimally, yielding a Nash welfare of $w + \log(1+2^x)$. So the approximation ratio achieved by the centralized solution is
\[
\frac{w+1}{w+\log(1+2^{2x})} < \frac{w+1}{w + 2x}
\]
which goes to $0$ as $x$ grows, for any constant $w$.

\subsection{Proof of Proposition~\ref{prop:u geq u star}}
\begin{proof}
If $|A_i| = 0$, then $u_i(A_i)=u^*_i(A_i)$ via simple algebra.
Else if $|A_i| > 0$, then
\begin{align*}
    u_i(A_i)
    &= \sum_{g \in A_i} \log\biggl(1 + \frac{v_i(g)}{e_i}\biggr)
    \geq \log \sum_{g \in A_i} \biggl(1 + \frac{v_i(g)}{e_i}\biggr)\\
    &= \log \biggl(|A_i| + \sum_{g \in A_i} \frac{v_i(g)}{e_i}\biggr) 
    \geq \log \biggl(1 + \sum_{g \in A_i} \frac{v_i(g)}{e_i}\biggr) 
    = u^*_i(A_i) \,.
\end{align*}\qed
\end{proof}

\subsection{Proof of Proposition~\ref{prop:u approx u star}}\label{apdx:proof u approx u star}

\begin{proof}
First, we rewrite the expression for $u_i$.
\begin{align*}
    u_i(g) 
    &= \log(e_i + v_i(g)) - \log(e_i) 
    = \log\biggl(1 + \frac{v_i(g)}{e_i}\biggr) \\
    u_i(A) 
    &= \sum_{g \in A_i} u_i(g) 
    = \sum_{g \in A_i} \log\biggl(1 + \frac{v_i(g)}{e_i}\biggr)
\end{align*}
If we substitute for $\log x = \frac{1}{\ln 2} \cdot \ln x$ its first order Taylor polynomial at 1, $P_1(x) = \frac{1}{\ln 2} (x - 1)$, we get the following:
\begin{align*}
    u_i(g) &\approx \frac{1}{\ln 2} \cdot \frac{v_i(g)}{e_i} \\
    u_i(A_i) &\approx \frac{1}{\ln 2} \cdot \frac{v_i(A_i)}{e_i} = \frac{1}{\ln 2} \cdot C_i
\end{align*}
Next, we rewrite the expression for $u^*_i$.
\begin{align*}
    u^*_i(A_i)
    &= \log\Bigl(e_i + \sum_{g \in A_i} v_i(g)\Bigr) - \log(e_i) \\
    &= \log\biggl(1 + \sum_{g \in A_i} \frac{v_i(g)}{e_i}\biggr) 
    = \log\biggl(1 + \frac{v_i(A_i)}{e_i}\biggr) 
    = \log(1 + C_i)
\end{align*}
Again, substituting in the Taylor polynomial, we have
\[u^*_i(A_i) \approx \frac{1}{\ln 2} \cdot C_i \]
Thus, $u$ is equivalent to $u^*$ at the first order level.

Now, we compare the two utility functions beyond the first order level.
By Taylor's theorem, for $f$ smooth, there exists for each $x$ some $b \in [1,x]$ such that
\begin{align*}
f(x) = f(1) + f'(1)(x-1)
+ \frac{f''(1)}{2!} (x-1)^2 + \frac{f^{(3)}(b)}{3!} (x-1)^3 \,.
\end{align*}
For $f(x) = \ln x$, this gives us
\[\ln x = x-1 - \frac{1}{2} (x-1)^2 + \frac{1}{3b^3} (x-1)^3 \,.\]
Using this identity in the bound for $u_i(A_i)$ gives us the following.
\begin{align*}
    u_i(g) \ln 2
    &= \ln\biggl(1 + \frac{v_i(g)}{e_i}\biggr) \\
    &= \frac{v_i(g)}{e_i} - \frac{1}{2} \biggl(\frac{v_i(g)}{e_i}\biggr)^2    + \frac{1}{3b_g^3} \biggl(\frac{v_i(g)}{e_i}\biggr)^3 \\
    &\leq \frac{v_i(g)}{e_i} - \frac{1}{2} \biggl(\frac{v_i(g)}{e_i}\biggr)^2 + \frac{1}{3} \biggl(\frac{v_i(g)}{e_i}\biggr)^3 \\
    u_i(A_i) \ln 2
    &= \sum_{g \in A_i} u_i(g) \ln 2 \\
    &
    \leq \frac{v_i(A_i)}{e_i} - \frac{1}{2e_i^2} \sum_{g \in A_i} v_i(g)^2 
    + \frac{1}{3e_i^3} \sum_{g \in A_i} v_i(g)^3
    \\
    &\leq \frac{v_i(A_i)}{e_i} + \frac{1}{3e_i^3} \cdot v_i(A_i)^3 \\
    &= C_i + \frac{C_i^3}{3}
\end{align*}

Finally, since $u_i(A_i) \geq u^*_i(A_i)$ by Proposition~\ref{prop:u geq u star}, the relative error of $u_i(A_i)$ as an approximation to $u^*_i(A_i)$ can be bounded as follows.
\begin{align*}
    \frac{u_i(A_i) - u^*_i(A_i)}{u^*_i(A_i)}
    &= \frac{u_i(A_i) \ln 2 - u^*_i(A_i) \ln 2}{u^*_i(A_i) \ln 2} \\
    &\leq \frac{C_i + \frac{C_i^3}{3} - \ln(1 + C_i)}{\ln(1 + C_i)}
    = \frac{C_i + \frac{C_i^3}{3}}{\ln(1 + C_i)} - 1
\end{align*}\qed
\end{proof}

\subsection{Proof of Theorem~\ref{thm:CEN counterexamples}}

\begin{proof}
    Consider a fair division instance with $n$ agents and $n$ goods, with $e_i = 1/x$ for all agents. Suppose $v_i(g_i) = \epsilon$ for all $i$, and let $v_i(g_{i-1}) = 1 - \epsilon$ for all $2 \leq i \leq n$, where $\epsilon = 0.001$. Suppose all other valuations are 0. Note that $\max_i v_i(G)/e_i = x$.

    The centralized solution will set $\CEN_i = \set{g_i}$, yielding a Nash welfare of $n \cdot \log(1/x + \epsilon)$. However, Nash welfare is maximized when $\OPT_1 = \emptyset$ and $\OPT_i = \set{g_{i-1}}$ for $i$ between 2 and $n-1$, and $\OPT_n = \set{g_{n-1}, g_n}$, yielding a Nash welfare of $\log(1/x) + \log(1/x + 1) + (n-2) \log(1/x + 1 - \epsilon)$.

    Next, note that
    $\log(1/x + 1 - \epsilon) - z \log(1/x + \epsilon) > 0$
    for all $x \geq 0.194$. Thus, for all
    \[n > \frac{\log(1/x + 1 - \epsilon) - \log(1/x)}{\log(1/x + 1 - \epsilon) - z \log(1/x + \epsilon)}\]
    we have $z \cdot n \log(1/x + \epsilon) < \log(1/x) + (n-1) \log(1/x + 1 - \epsilon)$ and subsequently
    \[
    z \cdot \NW\bigl(\CEN\bigr) < \NW\bigl(\OPT\bigr) 
    \]\qed
\end{proof}

\subsection{Proof of Theorem~\ref{thm:CPLI approx MNW}}
\begin{proof}
For every agent $i$ and good $g \in \OPT_i$ we have $e_i \leq v_i(g) \cdot \alpha \leq v_i(\OPT_i) \cdot \alpha$. Furthermore,
\begin{align*}
\log(v_i(\OPT_i) + e_i)
&\leq \log((1 + \alpha) \cdot v_i(\OPT_i)) 
= \log(1 + \alpha) + \log(v_i(\OPT_i)) \,.
\end{align*}
Summing across agents, we have 
\begin{align*}
    \NW\bigl(\OPT\bigr)
    &\leq n \cdot \log(1 + \alpha) + \sum_{i\in N} \log\bigl(v_i\bigl(\OPT_i\bigr)\bigr) \\
    &\leq n \cdot \log(1 + \alpha) + \sum_{i\in N} \log\bigl(v_i\bigl(\CEN_i\bigr)\bigr) \\
    &
    < n \cdot \log(1 + \alpha)
    + \sum_{i\in N} \log\bigl(v_i\bigl(\CEN_i\bigr) + e_i\bigr)
    \\
    &= n \cdot \log(1 + \alpha) + \NW\bigl(\CEN\bigr)
\end{align*}
where the second inequality follows from the fact that $\CEN$ is an allocation in 
$\mathbf{A}$ maximizing $\sum_{i \in N} \log(v_i(A_i))$, and the (final) equality follows from the definition of Nash welfare.\qed
\end{proof}

\subsection{Proof of Theorem~\ref{thm:inbetween}}

\begin{proof}
Without loss of generality, assume $e_1 \leq e_2 \leq \dots \leq e_n$. First, we calculate the probability that Nash welfare increases in the first step. Denote the first two agents to exchange goods by $i$ and $j$, with $i < j$. Note that neither agent $i$ nor $j$ is agent $1$ with probability $\frac{n-2}{n}$. For the rest of the proof, we assume that this is the case.

Nash welfare increases if $e_j > e_i + 1$. The probability of this happening is equal to the probability that two independent samples from the uniform distribution on $(e_1, M]$ differ by at least 1. This can be shown to be equal to $(M - e_1 - 1)^2 / (M - e_1)^2$ via a simple geometric probability argument.
The distribution of $e_1$ is $\beta(1,n,0,M)$. Using the CDF, we have that $e_1 \leq \frac{M}{n}$ with probability $1 - \bigl(1 - \frac{1}{n}\bigr)^n$. Thus, $e_j > e_i + 1$ with probability at least $(M - M/n - 1)^2 / (M - M/n)^2$ if $e_1 \leq \frac{M}{n}$.

Next, we show that Nash welfare decreases in the last step. Note that the last step involves some agent giving all their goods to agent $1$. In the best case (with respect to Nash welfare), that agent is agent $n$ and they are giving up only one good. This step reduces Nash welfare if $e_1 + n-1 > e_n$.

The distribution of $e_n$ is $\beta(n,1,0,M)$. Using the CDF, we have that $e_n < n - 1$ with probability at least \[\min\biggl(\biggl(\frac{n - 1}{M}\biggr)^n, 1\biggr) \,.\]
%
Multiplying all the probabilities gives the desired bound.\qed
\end{proof}

\section{Proofs from Section~\ref{sec:fairness tradeoffs}}

\subsection{Proof of Lemma \ref{lemma:expected vals}}~\label{apdx:expected vals lemma}

\begin{proof}
We can rewrite the expectations as follows.
\begin{align*}
\E[X \mid Z \land (X \geq Y)]
= \int \E[X \mid Z \land (X \geq y)] \cdot f_Y(y) \, dy
\end{align*} 
\begin{align*}
\E[X \mid Z \land (X \leq Y)]
= \int \E[X \mid Z \land (X \leq y)] \cdot f_Y(y) \, dy
\end{align*}
Note that for all $y$, we have
\begin{align*}
   \E[X \mid Z \land (X \geq y)] 
   \geq{} \E[X \mid Z] 
   \geq{} \E[X \mid Z \land (X \leq y)] \,.
\end{align*}
Multiplying by $f_Y(y)$ and integrating with respect to $y$, we get our desired result.\qed
\end{proof}

\subsection{Proof of Corollary~\ref{cor:EREF uniform}}\label{apdx:EREF uniform corollary proof}

\begin{proof}

For notational simplicity, use $A$ to denote $\DEC$. The proof of Theorem~\ref{thm:1wayEF expectation} proceeded by showing that $\Pr(g\in A_i) \geq \Pr(g \in A_j)$ and $\E[v_i(g) \mid g \in A_i] \geq \E[v_i(g) \mid g \in A_j]$ when $e_i \leq e_j$ and valuations for each good are drawn independently from a distribution $\mathcal{D}_g$ specific to that good. Multiplying these probabilities and summing them over all goods yields
\begin{align}
    \mathrel{\phantom{\iff}} &\sum_{g \in G} \E[v_i(g) \mid g \in A_i] \cdot Pr(g\in A_i) \geq \sum_{g \in G} \E[v_i(g) \mid g \in A_j] \cdot \Pr(g \in A_j) \label{math:exp-EREF-summary1}
    \\ \Longrightarrow &\E[v_i(A_i)] \geq \E[v_i(A_j)]\label{math:exp-EREF-summary2}
\end{align}

To prove this corollary, we will show that when $\mathcal{D}_g$ is the uniform distribution over $[0, M_g]$, we get the stronger inequality 
\begin{align*}
    \Pr(g\in A_i) \geq \frac{e_j}{e_i}\Pr(g \in A_j)
\end{align*}
which yields 
$\E[v_i(A_i)] \geq \frac{e_j}{e_i} \E[v_i(A_j)]$
via the same logic used to transition from Eq.~\ref{math:exp-EREF-summary1} to Eq.~\ref{math:exp-EREF-summary2}.

As in the proof of Theorem~\ref{thm:1wayEF expectation}, let $F$ be the CDF of $\mathcal{D}_g$. Suppose $\mathcal{D}_g$ is the uniform distribution over $[0, M_g]$. If $x \leq \frac{M_g e_i}{e_j}$, we have the following:

\begin{align*}
\Pr(g \in A_i \mid v_i(g) = x)
&= \prod_{k \neq i} F \biggl(x\frac{e_k}{e_i}\biggr) 
= F\biggl(x\frac{e_j}{e_i}\biggr) \cdot \prod_{k \neq i,j} F\biggl(x\frac{e_k}{e_i}\biggr) \\
&= x\frac{e_j}{M_g e_i} \cdot \prod_{k \neq i,j} F\biggl(x\frac{e_k}{e_i}\biggr) 
= \biggl(\frac{e_j}{e_i}\biggr)^2 x\frac{e_i}{M_g e_j} \cdot \prod_{k \neq i,j} F\biggl(x\frac{e_k}{e_i}\biggr) \\
&= \biggl(\frac{e_j}{e_i}\biggr)^2 F\biggl(x\frac{e_i}{e_j}\biggr) \cdot \prod_{k \neq i,j} F\biggl(x\frac{e_k}{e_i}\biggr) \\
&\geq \biggl(\frac{e_j}{e_i}\biggr)^2 F\biggl(x\frac{e_i}{e_j}\biggr) \cdot \prod_{k \neq i,j} F\biggl(x\frac{e_k}{e_j}\biggr) \\
&= \biggl(\frac{e_j}{e_i}\biggr)^2 \prod_{k \neq j} F\biggl(x\frac{e_k}{e_j}\biggr) = \biggl(\frac{e_j}{e_i}\biggr)^2 \Pr(g \in A_j \mid v_j(g) = x) \\
&\geq \frac{e_j}{e_i} \Pr(g \in A_j \mid v_j(g) = x)
\end{align*}
If $x > \frac{M_g e_i}{e_j}$, we have the following.
\begin{align*}
\Pr(g \in A_i \mid v_i(g) = x)
&= \prod_{k \neq i} F \biggl(x\frac{e_k}{e_i}\biggr) 
= F\biggl(x\frac{e_j}{e_i}\biggr) \prod_{k \neq i,j} F\biggl(x\frac{e_k}{e_i}\biggr) \\
&= 1 \prod_{k \neq i,j} F\biggl(x\frac{e_k}{e_i}\biggr) 
\geq \frac{x}{M_g} \prod_{k \neq i,j} F\biggl(x\frac{e_k}{e_i}\biggr) \\
&= \frac{e_j}{e_i} \cdot x\frac{e_i}{M_g e_j} \prod_{k \neq i,j} F\biggl(x\frac{e_k}{e_i}\biggr) \\
&= \frac{e_j}{e_i} \cdot F\biggl(x\frac{e_i}{e_j}\biggr) \prod_{k \neq i,j} F\biggl(x\frac{e_k}{e_i}\biggr) \\
&\geq \frac{e_j}{e_i} \cdot F\biggl(x\frac{e_i}{e_j}\biggr) \prod_{k \neq i,j} F\biggl(x\frac{e_k}{e_j}\biggr) 
= \frac{e_j}{e_i} \prod_{k \neq j} F\biggl(x\frac{e_k}{e_j}\biggr) \\
&= \frac{e_j}{e_i} \Pr(g \in A_j \mid v_j(g) = x)
\end{align*}
Via the law of total probability, we can write $\Pr(g \in A_i)$ as an integral, yielding the following:
\begin{align*}
    \Pr(g\in A_i) &= \int_0^{M_g} \Pr(g \in A_i \mid v_i(g) = x) \cdot \frac{1}{M_g} \, dx \\
    &\geq \int_0^{M_g} \frac{e_j}{e_i} \Pr(g \in A_j \mid v_j(g) = x) \cdot \frac{1}{M_g} \, dx \\
    &= \frac{e_j}{e_i} \int_0^{M_g} \Pr(g \in A_j \mid v_j(g) = x) \cdot \frac{1}{M_g} \, dx \\
    &= \frac{e_j}{e_i} \Pr(g\in A_j)
\end{align*}\qed

\end{proof}

\subsection{Proof of Theorem~\ref{thm:1wayEF probability} (\EREF{} with high probability)}
\begin{proof}
For notational simplicity, use $A$ to denote $\DEC$.
We want to show that
\[\Pr(v_i(A_i) \geq v_i(A_j)) \geq \Biggl(1 - \frac{\prod e_k}{n e_j^n}\Biggr)^m \,.\]
First, we'll show that we can focus specifically on the case of one good. We have
\begin{align*}
&\phanrel \Pr(v_i(A_i) \geq v_i(A_j)) \\
&= \Pr\Bigl(\sum_{g \in G} v_i(g) \cdot \ind(g \in A_i) \geq \sum_{g \in G} v_i(g) \cdot \ind(g \in A_j) \Bigr) \\
&\geq \prod_{g \in G} \Pr\bigl(v_i(g) \cdot \ind(g \in A_i) \geq v_i(g) \cdot \ind(g \in A_j)\bigr) \,,
\end{align*}
so to prove the theorem, it suffices to show
\[
\Pr\bigl(v_i(g) \cdot \ind(g \in A_i) \geq v_i(g) \cdot \ind(g \in A_j)\bigr) \geq 1 - \frac{\prod e_k}{n e_j^n}
\]
for an arbitrary good $g$.

We have the following.
\begin{align*}
    &\phanrel \Pr\bigl(v_i(g) \cdot \ind(g \in A_i) \geq v_i(g) \cdot \ind(g \in A_j)\bigr) \\
    &= \Pr\bigl(v_i(g) \cdot 1 \geq v_i(g) \cdot 0 \mathrel{\bigm|} g \in A_i\bigr) \cdot \Pr(g \in A_i) \\
    &\quad + \Pr\bigl(v_i(g) \cdot 0 \geq v_i(g) \cdot 1 \mathrel{\bigm|} g \in A_j\bigr) \cdot \Pr(g \in A_j) \\
    &\quad + \Pr\bigl(v_i(g) \cdot 0 \geq v_i(g) \cdot 0 \mathrel{\bigm|} g \notin A_i \cup A_j\bigr) \cdot \Pr(g \notin A_i \cup A_j) \\
    &= \Pr(g \in A_i) + \Pr(g \notin A_i \cup A_j) \\ 
    &= \Pr(g \notin A_j)
\end{align*}
By the definition of the decentralized mechanism, $g$ is put into $A_j$ if and only if 
\begin{align}\label{math:alloc rule v2}
\frac{v_j(g)}{e_j} \geq \frac{v_k(g)}{e_k} 
\end{align}
for all agents $k \ne j$. Let $F(x) = \frac{x}{M_g}$ be the CDF of $\mathcal{U}_{[0, M_g]}$. Then, for all $x \in [0, M_g]$, we have
\begin{align*}
\Pr(g \in A_j \mid v_j(g) = x) &= \Pr\biggl(\bigwedge_{k \neq j} v_k(g) \leq x \frac{e_k}{e_j} \biggr) \\
&= \prod_{k \neq j} F\biggl(x\frac{e_k}{e_j}\biggr) \\
&\leq \prod_{k\neq j} x\frac{e_k}{M_g e_j} \\
&= \biggl(\frac{x}{M_g e_j}\biggr)^{n-1} \Bigl(\prod_{k \neq j} e_k\Bigr) \,,
\end{align*}
where the first equality follows from the definition of the decentralized mechanism and from rearranging Eq.~\ref{math:alloc rule v2} and the second equality follows from the independence of the $v_k(g)$ terms.

Let $f = \frac{1}{M_g}$ be the PDF of $\mathcal{U}_{[0, M_g]}$.
Via the law of total probability, we can write $\Pr(g \in A_j)$ as an integral.
\begin{align*}
    \Pr(g\in A_j)
    &= \int_0^{M_g} \Pr(g \in A_j \mid v_j(g) = x) f(x) \, dx \\
    &\leq \int_0^{M_g} \biggl(\frac{x}{M_g e_j}\biggr)^{n-1} \Bigl(\prod_{k \neq j} e_k\Bigr) \frac{1}{M_g} \, dx \\
    &= \frac{\prod_{k \neq j} e_k}{M_g^n e_j^{n-1}} \int_0^{M_g} x^{n-1} \, dx \\
    &= \frac{\prod e_k}{M_g^n e_j^n} \cdot \frac{1}{n} x^n \biggr|_0^{M_g} \\
    &= \frac{\prod e_k}{n e_j^n}
\end{align*}
Then, we have
\[\Pr(g \notin A_j) \geq 1 - \frac{\prod e_k}{n e_j^n} \,,\]
and we are done.\qed
\end{proof}

\section{Additional Simulations}
\begin{figure}[bhpt]
    \centering
    \includegraphics[width=0.49\linewidth]{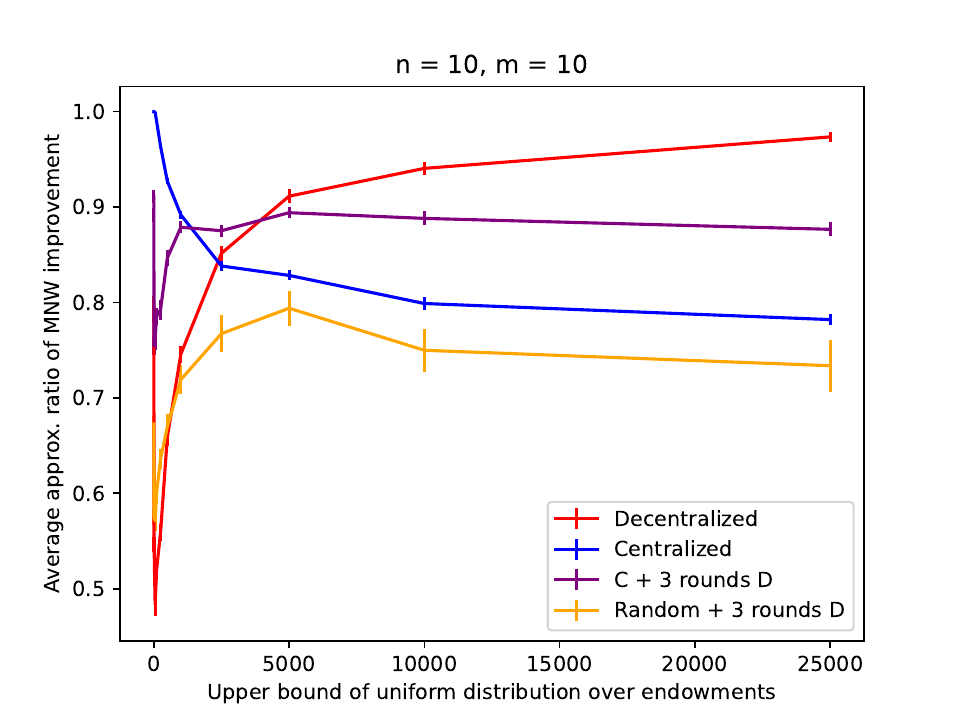}
    \includegraphics[width=0.49\linewidth]{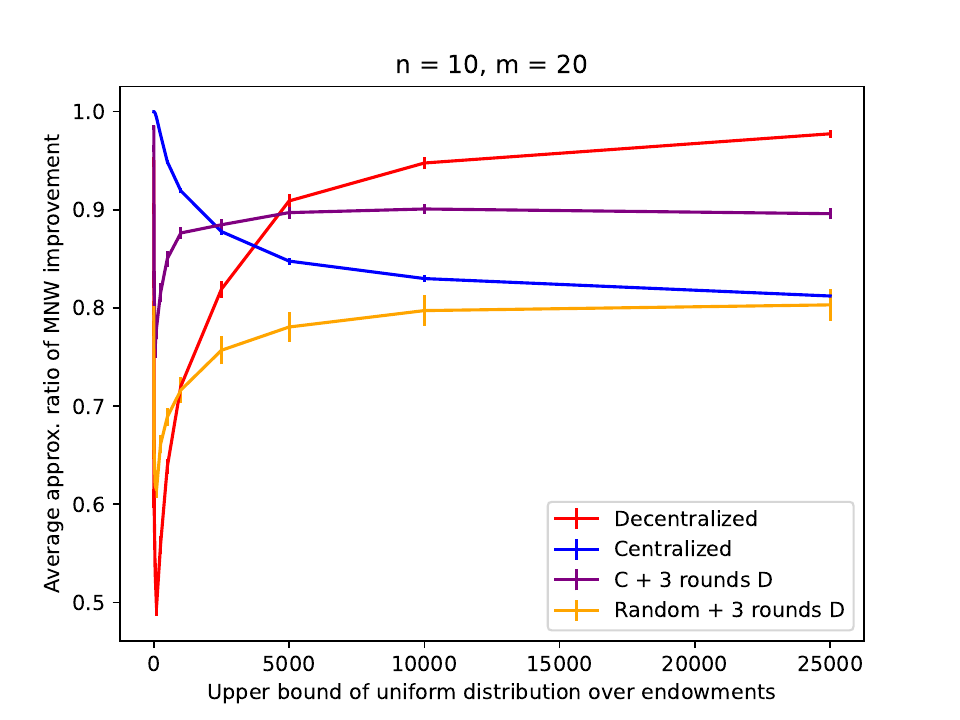}
    \\
    \includegraphics[width=0.49\linewidth]{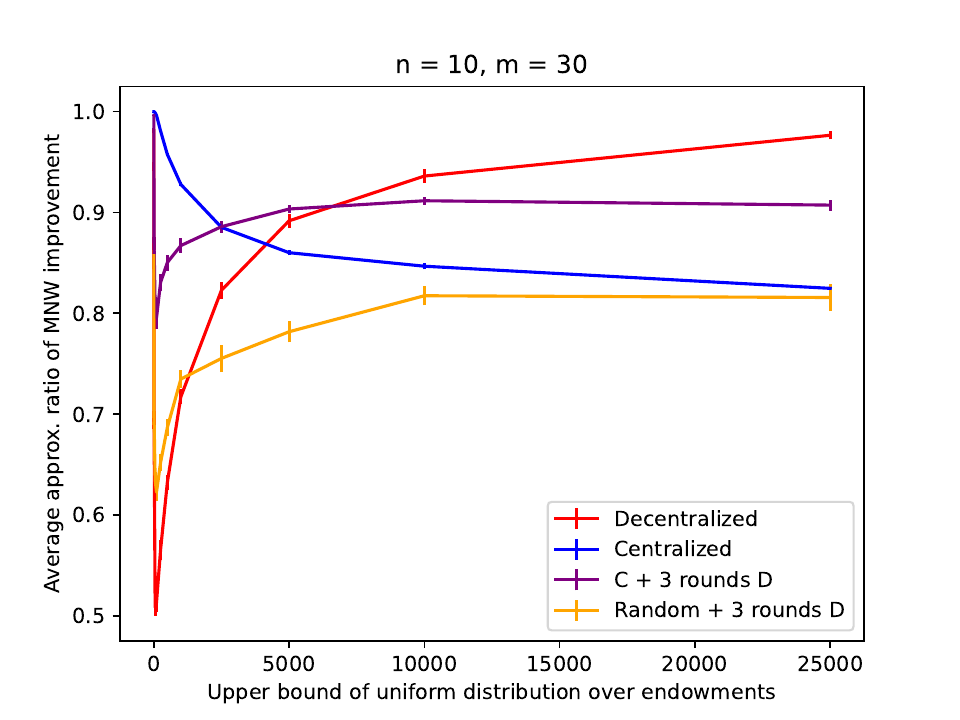}
    \includegraphics[width=0.49\linewidth]{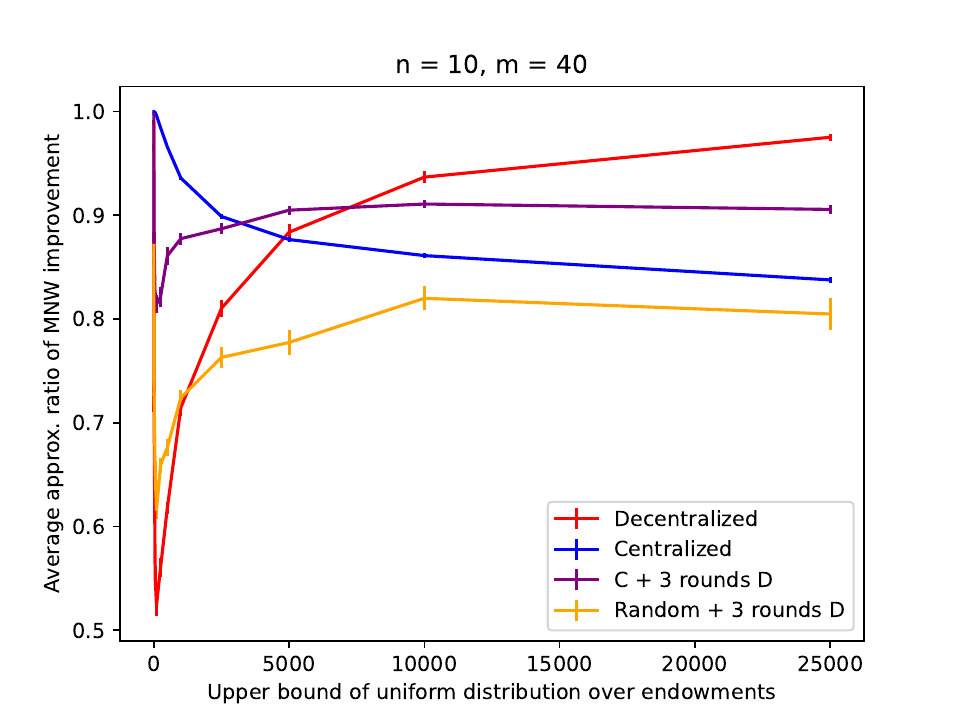}
    \\
    \includegraphics[width=0.49\linewidth]{n10m50}
    \caption{Approximation ratios for $n=10$ and $m\in \set{10,20,30,40,50}$.}
    \label{fig:supp-n=10}
\end{figure}

\begin{figure}[]
    \centering
    \includegraphics[width=0.49\linewidth]{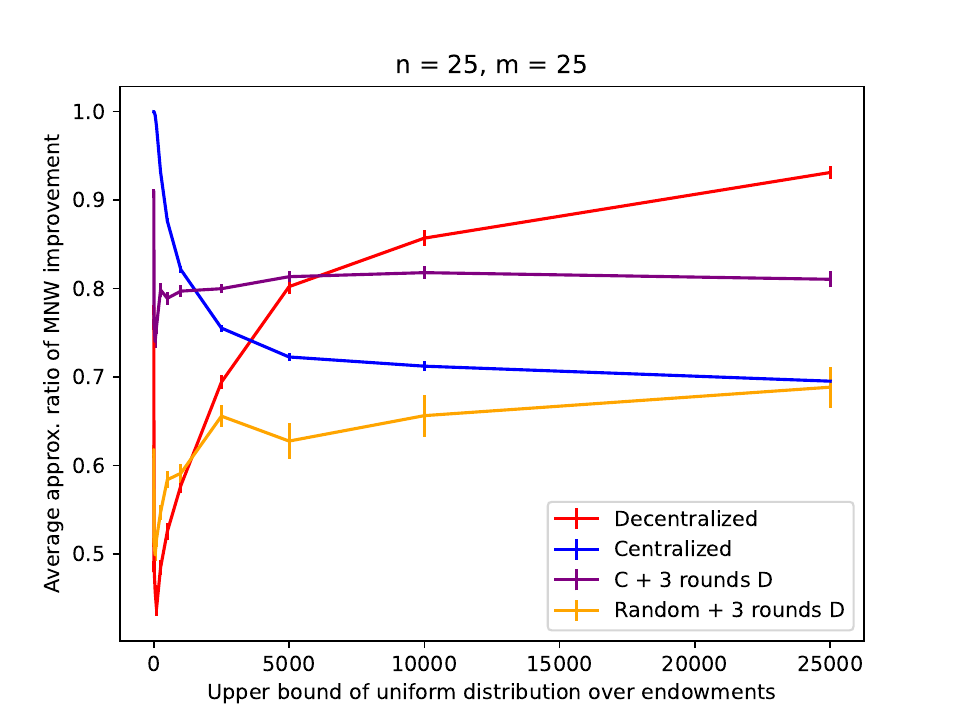}
    \includegraphics[width=0.49\linewidth]{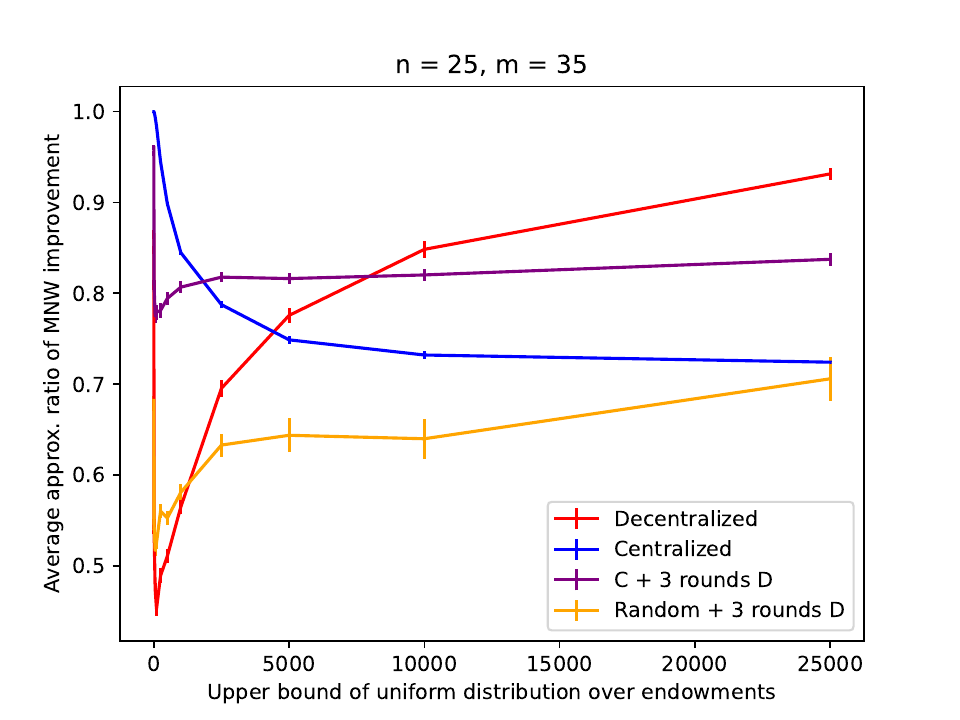}
    \\
    \includegraphics[width=0.49\linewidth]{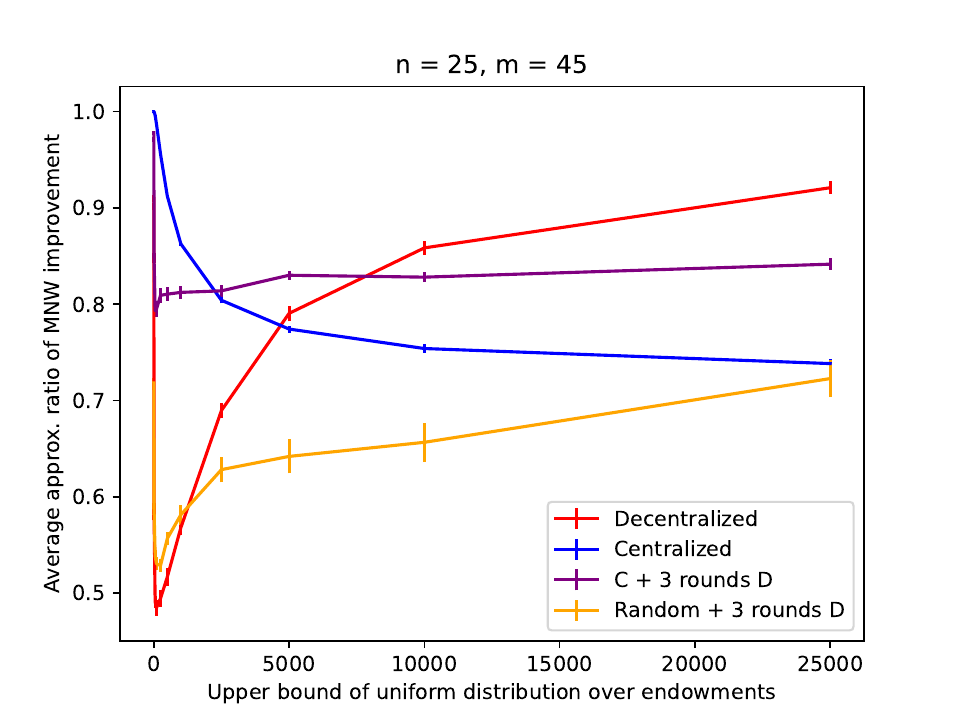}
    \includegraphics[width=0.49\linewidth]{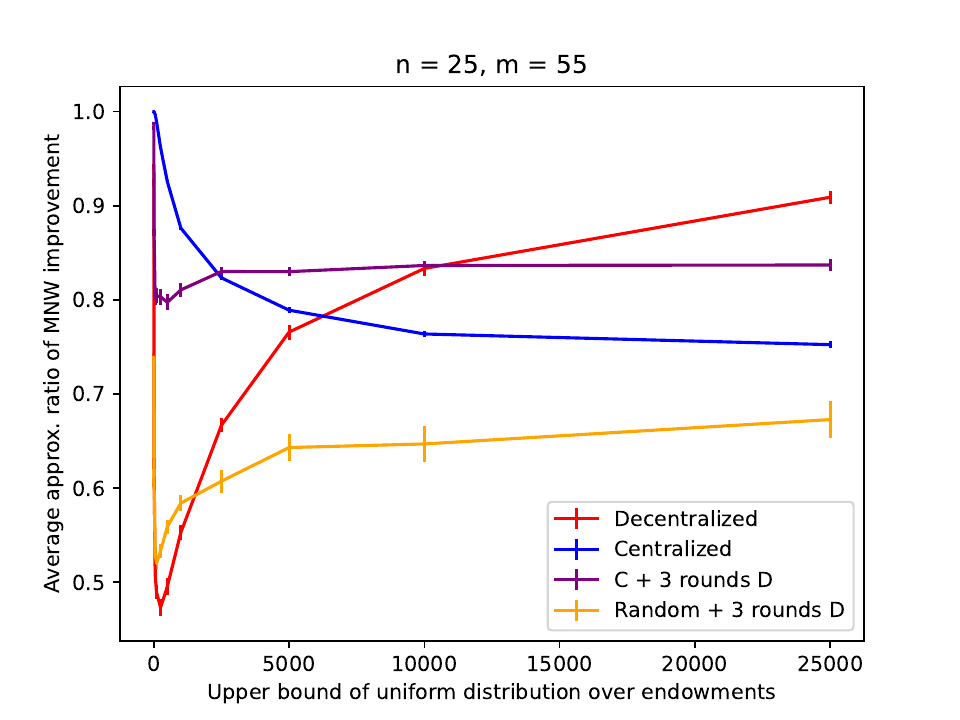}
    \caption{Approximation ratios for $n=25$ and $m\in \set{25,35,45,55}$.}
    \label{fig:supp-n=25}
\end{figure}

\begin{figure}[]
    \centering
    \includegraphics[width=0.49\linewidth]{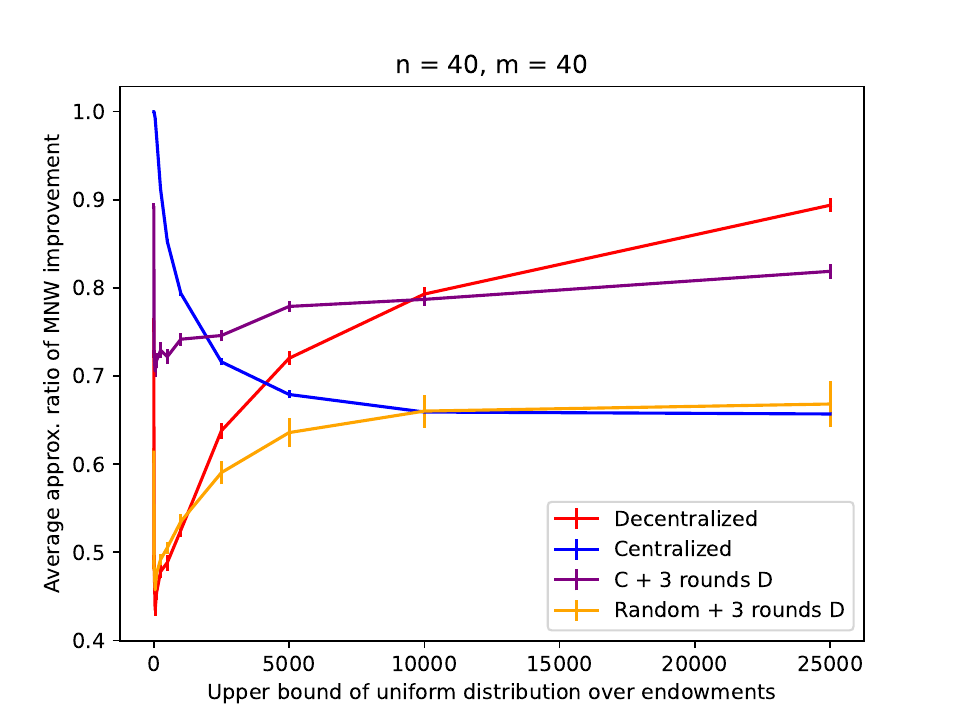}
    \includegraphics[width=0.49\linewidth]{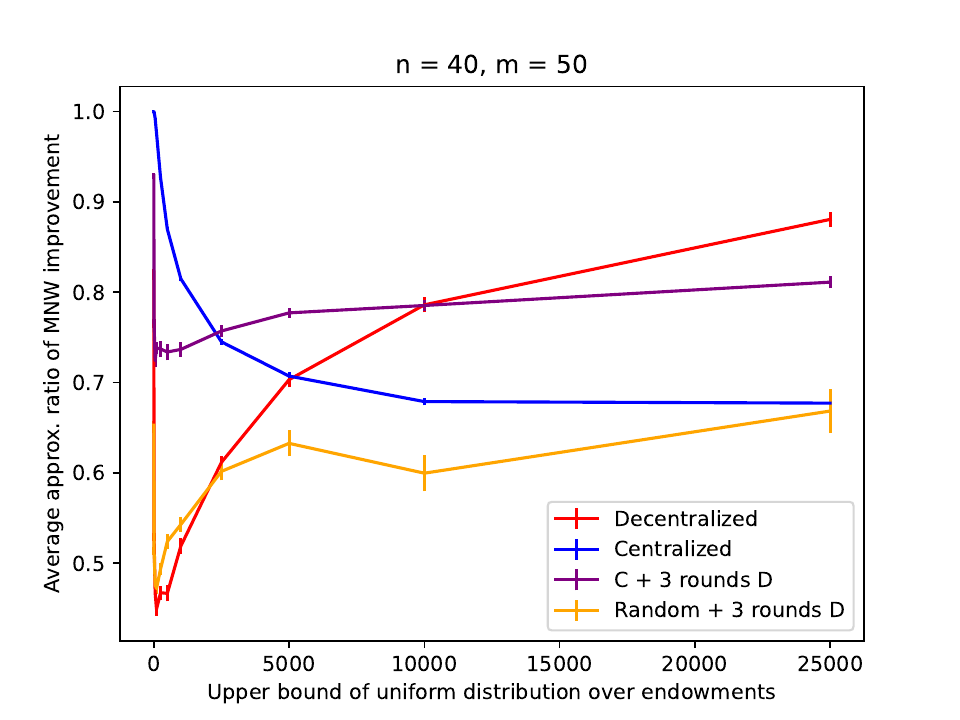}
    \\
    \includegraphics[width=0.49\linewidth]{n40m60}
    \includegraphics[width=0.49\linewidth]{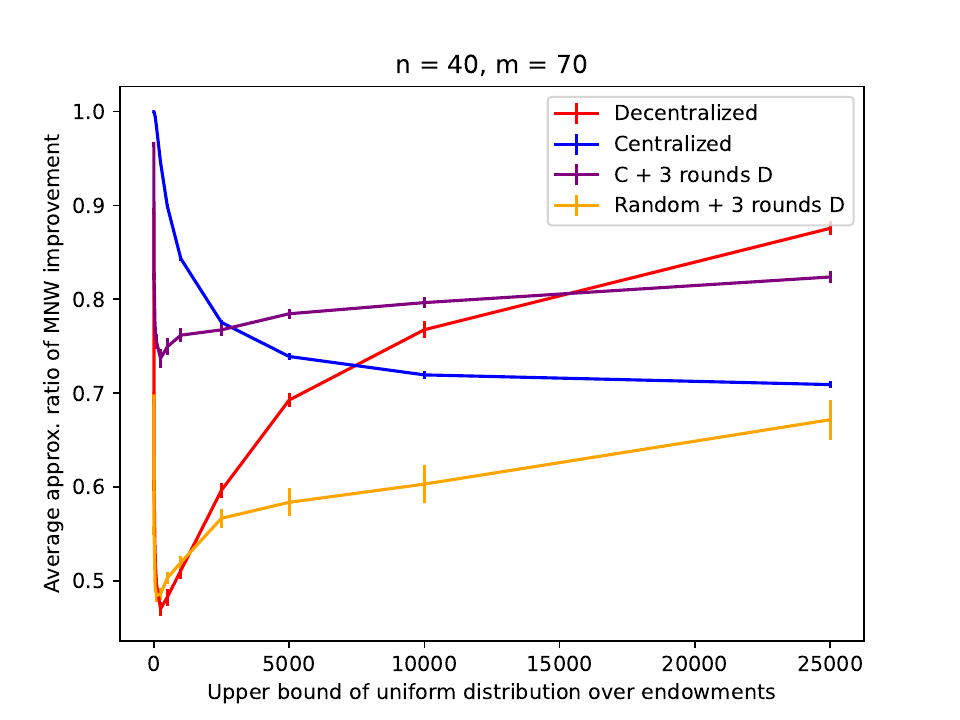}
    \\
    \includegraphics[width=0.49\linewidth]{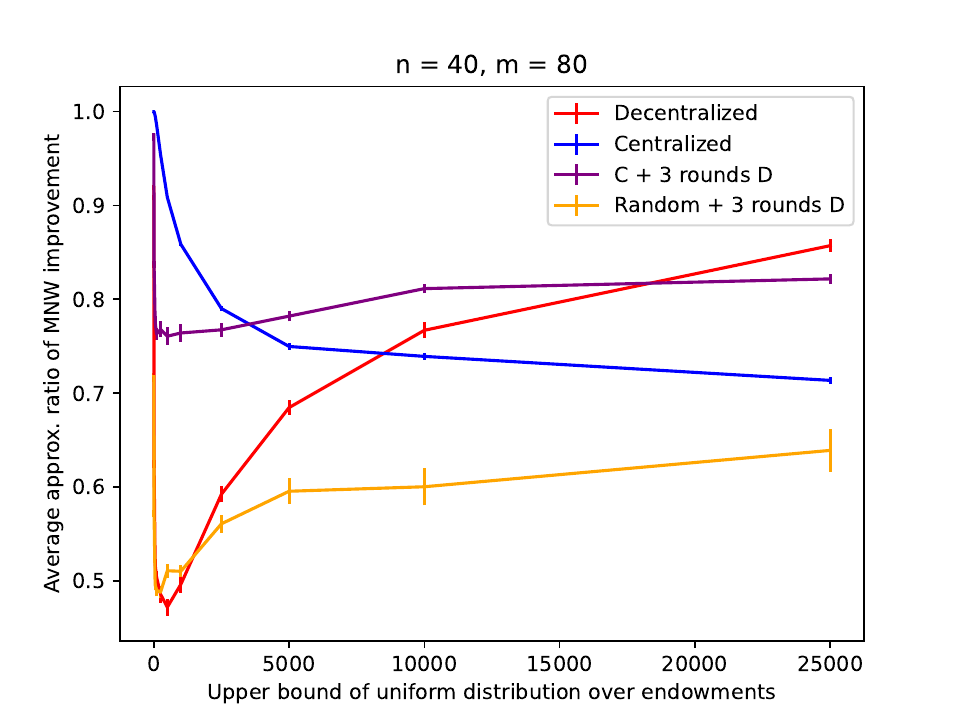}
    \caption{Approximation ratios for $n=40$ and $m\in \set{40,50,60,70,80}$.}
    \label{fig:supp-n=40}
\end{figure}

For a description of our simulation setup, see Section~\ref{sec:exps}. In Figures~\ref{fig:supp-n=10}--\ref{fig:supp-n=40}, we show the approximation ratios of various allocative schemes as a function of disparity in endowments for various values of $n$ and $m$. In all plots, tick marks represent standard error.

The wider selection of problem sizes on display here yield several additional observations about how $n$ and $m$ affect the approximation ratios of the allocative schemes. First, for a fixed value of $n$, changing $m$ does not appear to affect the quality of the decentralized solution, but it does appear to affect the quality of the centralized solution. In particular, increasing $m$ increases the quality of the centralized solution. Second, increasing the value of $n$ negatively affects the quality of both solutions.

Interestingly, when all endowments are exactly equal (when they are all $1$), the decentralized solution performs well, but then immediately drops when endowments are either $1$ or $2$. Since the decentralized solution simply finds an allocation maximizing utilitarian social welfare when all endowments are equal, the fact that it performs well when here may simply be a quirk of how we generated our random valuations: it could be the case that when valuations are random and evenly distributed, the allocation maximizing utilitarian social welfare also just happens to do a fairly good job of maximizing Nash welfare. In any case, after dipping to around a 50\% approximation ratio, the quality of the decentralized solution bounces back and eventually overtakes the centralized solution, which features a non-increasing approximation ratio as endowment disparity increases.

\end{document}